\documentclass[12pt,prd,showpacs,tightenlines,nofootinbib]{revtex4}
\usepackage{bm}
\usepackage{graphics}
\usepackage{rotating}
\usepackage{epsfig}

\begin{document}

\begin{flushright}
Preprint SSU-HEP-04/09\\
Samara State University\\
version 1
\end{flushright}

\title{2S HYPERFINE SPLITTING OF MUONIC HYDROGEN}

\author{A.P.Martynenko\footnote{mart@ssu.samara.ru}}

\address{Samara State University, Theoretical Physics
Department, Pavlov Street 1, 443011 Samara, Russia}

\begin{abstract}
Corrections of orders $\alpha^5$, $\alpha^6$ are calculated in the hyperfine
splitting of the $2S$ state in the muonic hydrogen. The nuclear structure
effects are taken into account in the one- and two-loop
Feynman amplitudes by means of the proton electromagnetic form factors.
Total numerical value of the $2S$ state hyperfine splitting 22.8148 meV
in the $\mu p$ can be considered as reliable estimation for the corresponding
experiment with the accuracy $10^{-5}$. The value of the Sternheim's
hyperfine splitting interval $[8\Delta E^{HFS}(2S)-\Delta E^{HFS}(1S)]$ is
obtained with the accuracy $10^{-6}$.
\end{abstract}


\maketitle

\immediate\write16{<<WARNING: LINEDRAW macros work with emTeX-dvivers
                    and other drivers supporting emTeX \special's
                    (dviscr, dvihplj, dvidot, dvips, dviwin, etc.) >>}

\newdimen\Lengthunit       \Lengthunit  = 1.5cm
\newcount\Nhalfperiods     \Nhalfperiods= 9
\newcount\magnitude        \magnitude = 1000

\catcode`\*=11
\newdimen\L*   \newdimen\d*   \newdimen\d**
\newdimen\dm*  \newdimen\dd*  \newdimen\dt*
\newdimen\a*   \newdimen\b*   \newdimen\c*
\newdimen\a**  \newdimen\b**
\newdimen\xL*  \newdimen\yL*
\newdimen\rx*  \newdimen\ry*
\newdimen\tmp* \newdimen\linwid*

\newcount\k*   \newcount\l*   \newcount\m*
\newcount\k**  \newcount\l**  \newcount\m**
\newcount\n*   \newcount\dn*  \newcount\r*
\newcount\N*   \newcount\*one \newcount\*two  \*one=1 \*two=2
\newcount\*ths \*ths=1000
\newcount\angle*  \newcount\q*  \newcount\q**
\newcount\angle** \angle**=0
\newcount\sc*     \sc*=0

\newtoks\cos*  \cos*={1}
\newtoks\sin*  \sin*={0}

\catcode`\[=13

\def\rotate(#1){\advance\angle**#1\angle*=\angle**
\q**=\angle*\ifnum\q**<0\q**=-\q**\fi
\ifnum\q**>360\q*=\angle*\divide\q*360\multiply\q*360\advance\angle*-\q*\fi
\ifnum\angle*<0\advance\angle*360\fi\q**=\angle*\divide\q**90\q**=\q**
\def\sgcos*{+}\def\sgsin*{+}\relax
\ifcase\q**\or
 \def\sgcos*{-}\def\sgsin*{+}\or
 \def\sgcos*{-}\def\sgsin*{-}\or
 \def\sgcos*{+}\def\sgsin*{-}\else\fi
\q*=\q**
\multiply\q*90\advance\angle*-\q*
\ifnum\angle*>45\sc*=1\angle*=-\angle*\advance\angle*90\else\sc*=0\fi
\def[##1,##2]{\ifnum\sc*=0\relax
\edef\cs*{\sgcos*.##1}\edef\sn*{\sgsin*.##2}\ifcase\q**\or
 \edef\cs*{\sgcos*.##2}\edef\sn*{\sgsin*.##1}\or
 \edef\cs*{\sgcos*.##1}\edef\sn*{\sgsin*.##2}\or
 \edef\cs*{\sgcos*.##2}\edef\sn*{\sgsin*.##1}\else\fi\else
\edef\cs*{\sgcos*.##2}\edef\sn*{\sgsin*.##1}\ifcase\q**\or
 \edef\cs*{\sgcos*.##1}\edef\sn*{\sgsin*.##2}\or
 \edef\cs*{\sgcos*.##2}\edef\sn*{\sgsin*.##1}\or
 \edef\cs*{\sgcos*.##1}\edef\sn*{\sgsin*.##2}\else\fi\fi
\cos*={\cs*}\sin*={\sn*}\global\edef\gcos*{\cs*}\global\edef\gsin*{\sn*}}\relax
\ifcase\angle*[9999,0]\or
[999,017]\or[999,034]\or[998,052]\or[997,069]\or[996,087]\or
[994,104]\or[992,121]\or[990,139]\or[987,156]\or[984,173]\or
[981,190]\or[978,207]\or[974,224]\or[970,241]\or[965,258]\or
[961,275]\or[956,292]\or[951,309]\or[945,325]\or[939,342]\or
[933,358]\or[927,374]\or[920,390]\or[913,406]\or[906,422]\or
[898,438]\or[891,453]\or[882,469]\or[874,484]\or[866,499]\or
[857,515]\or[848,529]\or[838,544]\or[829,559]\or[819,573]\or
[809,587]\or[798,601]\or[788,615]\or[777,629]\or[766,642]\or
[754,656]\or[743,669]\or[731,681]\or[719,694]\or[707,707]\or
\else[9999,0]\fi}

\catcode`\[=12

\def\GRAPH(hsize=#1)#2{\hbox to #1\Lengthunit{#2\hss}}

\def\Linewidth#1{\global\linwid*=#1\relax
\global\divide\linwid*10\global\multiply\linwid*\mag
\global\divide\linwid*100\special{em:linewidth \the\linwid*}}

\Linewidth{.4pt}
\def\sm*{\special{em:moveto}}
\def\sl*{\special{em:lineto}}
\let\moveto=\sm*
\let\lineto=\sl*
\newbox\spm*   \newbox\spl*
\setbox\spm*\hbox{\sm*}
\setbox\spl*\hbox{\sl*}

\def\mov#1(#2,#3)#4{\rlap{\L*=#1\Lengthunit
\xL*=#2\L* \yL*=#3\L*
\xL*=\xscale\xL* \yL*=\yscale\yL*
\rx* \the\cos*\xL* \tmp* \the\sin*\yL* \advance\rx*-\tmp*
\ry* \the\cos*\yL* \tmp* \the\sin*\xL* \advance\ry*\tmp*
\kern\rx*\raise\ry*\hbox{#4}}}

\def\rmov*(#1,#2)#3{\rlap{\xL*=#1\yL*=#2\relax
\rx* \the\cos*\xL* \tmp* \the\sin*\yL* \advance\rx*-\tmp*
\ry* \the\cos*\yL* \tmp* \the\sin*\xL* \advance\ry*\tmp*
\kern\rx*\raise\ry*\hbox{#3}}}

\def\lin#1(#2,#3){\rlap{\sm*\mov#1(#2,#3){\sl*}}}

\def\arr*(#1,#2,#3){\rmov*(#1\dd*,#1\dt*){\sm*
\rmov*(#2\dd*,#2\dt*){\rmov*(#3\dt*,-#3\dd*){\sl*}}\sm*
\rmov*(#2\dd*,#2\dt*){\rmov*(-#3\dt*,#3\dd*){\sl*}}}}

\def\arrow#1(#2,#3){\rlap{\lin#1(#2,#3)\mov#1(#2,#3){\relax
\d**=-.012\Lengthunit\dd*=#2\d**\dt*=#3\d**
\arr*(1,10,4)\arr*(3,8,4)\arr*(4.8,4.2,3)}}}

\def\arrlin#1(#2,#3){\rlap{\L*=#1\Lengthunit\L*=.5\L*
\lin#1(#2,#3)\rmov*(#2\L*,#3\L*){\arrow.1(#2,#3)}}}

\def\dasharrow#1(#2,#3){\rlap{{\Lengthunit=0.9\Lengthunit
\dashlin#1(#2,#3)\mov#1(#2,#3){\sm*}}\mov#1(#2,#3){\sl*
\d**=-.012\Lengthunit\dd*=#2\d**\dt*=#3\d**
\arr*(1,10,4)\arr*(3,8,4)\arr*(4.8,4.2,3)}}}

\def\clap#1{\hbox to 0pt{\hss #1\hss}}

\def\ind(#1,#2)#3{\rlap{\L*=.1\Lengthunit
\xL*=#1\L* \yL*=#2\L*
\rx* \the\cos*\xL* \tmp* \the\sin*\yL* \advance\rx*-\tmp*
\ry* \the\cos*\yL* \tmp* \the\sin*\xL* \advance\ry*\tmp*
\kern\rx*\raise\ry*\hbox{\lower2pt\clap{$#3$}}}}

\def\sh*(#1,#2)#3{\rlap{\dm*=\the\n*\d**
\xL*=\xscale\dm* \yL*=\yscale\dm* \xL*=#1\xL* \yL*=#2\yL*
\rx* \the\cos*\xL* \tmp* \the\sin*\yL* \advance\rx*-\tmp*
\ry* \the\cos*\yL* \tmp* \the\sin*\xL* \advance\ry*\tmp*
\kern\rx*\raise\ry*\hbox{#3}}}

\def\calcnum*#1(#2,#3){\a*=1000sp\b*=1000sp\a*=#2\a*\b*=#3\b*
\ifdim\a*<0pt\a*-\a*\fi\ifdim\b*<0pt\b*-\b*\fi
\ifdim\a*>\b*\c*=.96\a*\advance\c*.4\b*
\else\c*=.96\b*\advance\c*.4\a*\fi
\k*\a*\multiply\k*\k*\l*\b*\multiply\l*\l*
\m*\k*\advance\m*\l*\n*\c*\r*\n*\multiply\n*\n*
\dn*\m*\advance\dn*-\n*\divide\dn*2\divide\dn*\r*
\advance\r*\dn*
\c*=\the\Nhalfperiods5sp\c*=#1\c*\ifdim\c*<0pt\c*-\c*\fi
\multiply\c*\r*\N*\c*\divide\N*10000}

\def\dashlin#1(#2,#3){\rlap{\calcnum*#1(#2,#3)\relax
\d**=#1\Lengthunit\ifdim\d**<0pt\d**-\d**\fi
\divide\N*2\multiply\N*2\advance\N*\*one
\divide\d**\N*\sm*\n*\*one\sh*(#2,#3){\sl*}\loop
\advance\n*\*one\sh*(#2,#3){\sm*}\advance\n*\*one
\sh*(#2,#3){\sl*}\ifnum\n*<\N*\repeat}}

\def\dashdotlin#1(#2,#3){\rlap{\calcnum*#1(#2,#3)\relax
\d**=#1\Lengthunit\ifdim\d**<0pt\d**-\d**\fi
\divide\N*2\multiply\N*2\advance\N*1\multiply\N*2\relax
\divide\d**\N*\sm*\n*\*two\sh*(#2,#3){\sl*}\loop
\advance\n*\*one\sh*(#2,#3){\kern-1.48pt\lower.5pt\hbox{\rm.}}\relax
\advance\n*\*one\sh*(#2,#3){\sm*}\advance\n*\*two
\sh*(#2,#3){\sl*}\ifnum\n*<\N*\repeat}}

\def\shl*(#1,#2)#3{\kern#1#3\lower#2#3\hbox{\unhcopy\spl*}}

\def\trianglin#1(#2,#3){\rlap{\toks0={#2}\toks1={#3}\calcnum*#1(#2,#3)\relax
\dd*=.57\Lengthunit\dd*=#1\dd*\divide\dd*\N*
\divide\dd*\*ths \multiply\dd*\magnitude
\d**=#1\Lengthunit\ifdim\d**<0pt\d**-\d**\fi
\multiply\N*2\divide\d**\N*\sm*\n*\*one\loop
\shl**{\dd*}\dd*-\dd*\advance\n*2\relax
\ifnum\n*<\N*\repeat\n*\N*\shl**{0pt}}}

\def\wavelin#1(#2,#3){\rlap{\toks0={#2}\toks1={#3}\calcnum*#1(#2,#3)\relax
\dd*=.23\Lengthunit\dd*=#1\dd*\divide\dd*\N*
\divide\dd*\*ths \multiply\dd*\magnitude
\d**=#1\Lengthunit\ifdim\d**<0pt\d**-\d**\fi
\multiply\N*4\divide\d**\N*\sm*\n*\*one\loop
\shl**{\dd*}\dt*=1.3\dd*\advance\n*\*one
\shl**{\dt*}\advance\n*\*one
\shl**{\dd*}\advance\n*\*two
\dd*-\dd*\ifnum\n*<\N*\repeat\n*\N*\shl**{0pt}}}

\def\w*lin(#1,#2){\rlap{\toks0={#1}\toks1={#2}\d**=\Lengthunit\dd*=-.12\d**
\divide\dd*\*ths \multiply\dd*\magnitude
\N*8\divide\d**\N*\sm*\n*\*one\loop
\shl**{\dd*}\dt*=1.3\dd*\advance\n*\*one
\shl**{\dt*}\advance\n*\*one
\shl**{\dd*}\advance\n*\*one
\shl**{0pt}\dd*-\dd*\advance\n*1\ifnum\n*<\N*\repeat}}

\def\l*arc(#1,#2)[#3][#4]{\rlap{\toks0={#1}\toks1={#2}\d**=\Lengthunit
\dd*=#3.037\d**\dd*=#4\dd*\dt*=#3.049\d**\dt*=#4\dt*\ifdim\d**>10mm\relax
\d**=.25\d**\n*\*one\shl**{-\dd*}\n*\*two\shl**{-\dt*}\n*3\relax
\shl**{-\dd*}\n*4\relax\shl**{0pt}\else
\ifdim\d**>5mm\d**=.5\d**\n*\*one\shl**{-\dt*}\n*\*two
\shl**{0pt}\else\n*\*one\shl**{0pt}\fi\fi}}

\def\d*arc(#1,#2)[#3][#4]{\rlap{\toks0={#1}\toks1={#2}\d**=\Lengthunit
\dd*=#3.037\d**\dd*=#4\dd*\d**=.25\d**\sm*\n*\*one\shl**{-\dd*}\relax
\n*3\relax\sh*(#1,#2){\xL*=\xscale\dd*\yL*=\yscale\dd*
\kern#2\xL*\lower#1\yL*\hbox{\sm*}}\n*4\relax\shl**{0pt}}}

\def\shl**#1{\c*=\the\n*\d**\d*=#1\relax
\a*=\the\toks0\c*\b*=\the\toks1\d*\advance\a*-\b*
\b*=\the\toks1\c*\d*=\the\toks0\d*\advance\b*\d*
\a*=\xscale\a*\b*=\yscale\b*
\rx* \the\cos*\a* \tmp* \the\sin*\b* \advance\rx*-\tmp*
\ry* \the\cos*\b* \tmp* \the\sin*\a* \advance\ry*\tmp*
\raise\ry*\rlap{\kern\rx*\unhcopy\spl*}}

\def\wlin*#1(#2,#3)[#4]{\rlap{\toks0={#2}\toks1={#3}\relax
\c*=#1\l*\c*\c*=.01\Lengthunit\m*\c*\divide\l*\m*
\c*=\the\Nhalfperiods5sp\multiply\c*\l*\N*\c*\divide\N*\*ths
\divide\N*2\multiply\N*2\advance\N*\*one
\dd*=.002\Lengthunit\dd*=#4\dd*\multiply\dd*\l*\divide\dd*\N*
\divide\dd*\*ths \multiply\dd*\magnitude
\d**=#1\multiply\N*4\divide\d**\N*\sm*\n*\*one\loop
\shl**{\dd*}\dt*=1.3\dd*\advance\n*\*one
\shl**{\dt*}\advance\n*\*one
\shl**{\dd*}\advance\n*\*two
\dd*-\dd*\ifnum\n*<\N*\repeat\n*\N*\shl**{0pt}}}

\def\wavebox#1{\setbox0\hbox{#1}\relax
\a*=\wd0\advance\a*14pt\b*=\ht0\advance\b*\dp0\advance\b*14pt\relax
\hbox{\kern9pt\relax
\rmov*(0pt,\ht0){\rmov*(-7pt,7pt){\wlin*\a*(1,0)[+]\wlin*\b*(0,-1)[-]}}\relax
\rmov*(\wd0,-\dp0){\rmov*(7pt,-7pt){\wlin*\a*(-1,0)[+]\wlin*\b*(0,1)[-]}}\relax
\box0\kern9pt}}

\def\rectangle#1(#2,#3){\relax
\lin#1(#2,0)\lin#1(0,#3)\mov#1(0,#3){\lin#1(#2,0)}\mov#1(#2,0){\lin#1(0,#3)}}

\def\dashrectangle#1(#2,#3){\dashlin#1(#2,0)\dashlin#1(0,#3)\relax
\mov#1(0,#3){\dashlin#1(#2,0)}\mov#1(#2,0){\dashlin#1(0,#3)}}

\def\waverectangle#1(#2,#3){\L*=#1\Lengthunit\a*=#2\L*\b*=#3\L*
\ifdim\a*<0pt\a*-\a*\def\x*{-1}\else\def\x*{1}\fi
\ifdim\b*<0pt\b*-\b*\def\y*{-1}\else\def\y*{1}\fi
\wlin*\a*(\x*,0)[-]\wlin*\b*(0,\y*)[+]\relax
\mov#1(0,#3){\wlin*\a*(\x*,0)[+]}\mov#1(#2,0){\wlin*\b*(0,\y*)[-]}}

\def\calcparab*{\ifnum\n*>\m*\k*\N*\advance\k*-\n*\else\k*\n*\fi
\a*=\the\k* sp\a*=10\a*\b*\dm*\advance\b*-\a*\k*\b*
\a*=\the\*ths\b*\divide\a*\l*\multiply\a*\k*
\divide\a*\l*\k*\*ths\r*\a*\advance\k*-\r*\dt*=\the\k*\L*}

\def\arcto#1(#2,#3)[#4]{\rlap{\toks0={#2}\toks1={#3}\calcnum*#1(#2,#3)\relax
\dm*=135sp\dm*=#1\dm*\d**=#1\Lengthunit\ifdim\dm*<0pt\dm*-\dm*\fi
\multiply\dm*\r*\a*=.3\dm*\a*=#4\a*\ifdim\a*<0pt\a*-\a*\fi
\advance\dm*\a*\N*\dm*\divide\N*10000\relax
\divide\N*2\multiply\N*2\advance\N*\*one
\L*=-.25\d**\L*=#4\L*\divide\d**\N*\divide\L*\*ths
\m*\N*\divide\m*2\dm*=\the\m*5sp\l*\dm*\sm*\n*\*one\loop
\calcparab*\shl**{-\dt*}\advance\n*1\ifnum\n*<\N*\repeat}}

\def\arrarcto#1(#2,#3)[#4]{\L*=#1\Lengthunit\L*=.54\L*
\arcto#1(#2,#3)[#4]\rmov*(#2\L*,#3\L*){\d*=.457\L*\d*=#4\d*\d**-\d*
\rmov*(#3\d**,#2\d*){\arrow.02(#2,#3)}}}

\def\dasharcto#1(#2,#3)[#4]{\rlap{\toks0={#2}\toks1={#3}\relax
\calcnum*#1(#2,#3)\dm*=\the\N*5sp\a*=.3\dm*\a*=#4\a*\ifdim\a*<0pt\a*-\a*\fi
\advance\dm*\a*\N*\dm*
\divide\N*20\multiply\N*2\advance\N*1\d**=#1\Lengthunit
\L*=-.25\d**\L*=#4\L*\divide\d**\N*\divide\L*\*ths
\m*\N*\divide\m*2\dm*=\the\m*5sp\l*\dm*
\sm*\n*\*one\loop\calcparab*
\shl**{-\dt*}\advance\n*1\ifnum\n*>\N*\else\calcparab*
\sh*(#2,#3){\xL*=#3\dt* \yL*=#2\dt*
\rx* \the\cos*\xL* \tmp* \the\sin*\yL* \advance\rx*\tmp*
\ry* \the\cos*\yL* \tmp* \the\sin*\xL* \advance\ry*-\tmp*
\kern\rx*\lower\ry*\hbox{\sm*}}\fi
\advance\n*1\ifnum\n*<\N*\repeat}}

\def\*shl*#1{\c*=\the\n*\d**\advance\c*#1\a**\d*\dt*\advance\d*#1\b**
\a*=\the\toks0\c*\b*=\the\toks1\d*\advance\a*-\b*
\b*=\the\toks1\c*\d*=\the\toks0\d*\advance\b*\d*
\rx* \the\cos*\a* \tmp* \the\sin*\b* \advance\rx*-\tmp*
\ry* \the\cos*\b* \tmp* \the\sin*\a* \advance\ry*\tmp*
\raise\ry*\rlap{\kern\rx*\unhcopy\spl*}}

\def\calcnormal*#1{\b**=10000sp\a**\b**\k*\n*\advance\k*-\m*
\multiply\a**\k*\divide\a**\m*\a**=#1\a**\ifdim\a**<0pt\a**-\a**\fi
\ifdim\a**>\b**\d*=.96\a**\advance\d*.4\b**
\else\d*=.96\b**\advance\d*.4\a**\fi
\d*=.01\d*\r*\d*\divide\a**\r*\divide\b**\r*
\ifnum\k*<0\a**-\a**\fi\d*=#1\d*\ifdim\d*<0pt\b**-\b**\fi
\k*\a**\a**=\the\k*\dd*\k*\b**\b**=\the\k*\dd*}

\def\wavearcto#1(#2,#3)[#4]{\rlap{\toks0={#2}\toks1={#3}\relax
\calcnum*#1(#2,#3)\c*=\the\N*5sp\a*=.4\c*\a*=#4\a*\ifdim\a*<0pt\a*-\a*\fi
\advance\c*\a*\N*\c*\divide\N*20\multiply\N*2\advance\N*-1\multiply\N*4\relax
\d**=#1\Lengthunit\dd*=.012\d**
\divide\dd*\*ths \multiply\dd*\magnitude
\ifdim\d**<0pt\d**-\d**\fi\L*=.25\d**
\divide\d**\N*\divide\dd*\N*\L*=#4\L*\divide\L*\*ths
\m*\N*\divide\m*2\dm*=\the\m*0sp\l*\dm*
\sm*\n*\*one\loop\calcnormal*{#4}\calcparab*
\*shl*{1}\advance\n*\*one\calcparab*
\*shl*{1.3}\advance\n*\*one\calcparab*
\*shl*{1}\advance\n*2\dd*-\dd*\ifnum\n*<\N*\repeat\n*\N*\shl**{0pt}}}

\def\triangarcto#1(#2,#3)[#4]{\rlap{\toks0={#2}\toks1={#3}\relax
\calcnum*#1(#2,#3)\c*=\the\N*5sp\a*=.4\c*\a*=#4\a*\ifdim\a*<0pt\a*-\a*\fi
\advance\c*\a*\N*\c*\divide\N*20\multiply\N*2\advance\N*-1\multiply\N*2\relax
\d**=#1\Lengthunit\dd*=.012\d**
\divide\dd*\*ths \multiply\dd*\magnitude
\ifdim\d**<0pt\d**-\d**\fi\L*=.25\d**
\divide\d**\N*\divide\dd*\N*\L*=#4\L*\divide\L*\*ths
\m*\N*\divide\m*2\dm*=\the\m*0sp\l*\dm*
\sm*\n*\*one\loop\calcnormal*{#4}\calcparab*
\*shl*{1}\advance\n*2\dd*-\dd*\ifnum\n*<\N*\repeat\n*\N*\shl**{0pt}}}

\def\hr*#1{\L*=\xscale\Lengthunit\ifnum
\angle**=0\clap{\vrule width#1\L* height.1pt}\else
\L*=#1\L*\L*=.5\L*\rmov*(-\L*,0pt){\sm*}\rmov*(\L*,0pt){\sl*}\fi}

\def\shade#1[#2]{\rlap{\Lengthunit=#1\Lengthunit
\special{em:linewidth .001pt}\relax
\mov(0,#2.05){\hr*{.994}}\mov(0,#2.1){\hr*{.980}}\relax
\mov(0,#2.15){\hr*{.953}}\mov(0,#2.2){\hr*{.916}}\relax
\mov(0,#2.25){\hr*{.867}}\mov(0,#2.3){\hr*{.798}}\relax
\mov(0,#2.35){\hr*{.715}}\mov(0,#2.4){\hr*{.603}}\relax
\mov(0,#2.45){\hr*{.435}}\special{em:linewidth \the\linwid*}}}

\def\dshade#1[#2]{\rlap{\special{em:linewidth .001pt}\relax
\Lengthunit=#1\Lengthunit\if#2-\def\t*{+}\else\def\t*{-}\fi
\mov(0,\t*.025){\relax
\mov(0,#2.05){\hr*{.995}}\mov(0,#2.1){\hr*{.988}}\relax
\mov(0,#2.15){\hr*{.969}}\mov(0,#2.2){\hr*{.937}}\relax
\mov(0,#2.25){\hr*{.893}}\mov(0,#2.3){\hr*{.836}}\relax
\mov(0,#2.35){\hr*{.760}}\mov(0,#2.4){\hr*{.662}}\relax
\mov(0,#2.45){\hr*{.531}}\mov(0,#2.5){\hr*{.320}}\relax
\special{em:linewidth \the\linwid*}}}}

\def\vdot{\rlap{\kern-1.9pt\lower1.8pt\hbox{$\scriptstyle\bullet$}}}
\def\vtimes{\rlap{\kern-3pt\lower1.8pt\hbox{$\scriptstyle\times$}}}
\def\vDot{\rlap{\kern-2.3pt\lower2.7pt\hbox{$\bullet$}}}
\def\vTimes{\rlap{\kern-3.6pt\lower2.4pt\hbox{$\times$}}}

\def\arc(#1)[#2,#3]{{\k*=#2\l*=#3\m*=\l*
\advance\m*-6\ifnum\k*>\l*\relax\else
{\rotate(#2)\mov(#1,0){\sm*}}\loop
\ifnum\k*<\m*\advance\k*5{\rotate(\k*)\mov(#1,0){\sl*}}\repeat
{\rotate(#3)\mov(#1,0){\sl*}}\fi}}

\def\dasharc(#1)[#2,#3]{{\k**=#2\n*=#3\advance\n*-1\advance\n*-\k**
\L*=1000sp\L*#1\L* \multiply\L*\n* \multiply\L*\Nhalfperiods
\divide\L*57\N*\L* \divide\N*2000\ifnum\N*=0\N*1\fi
\r*\n*  \divide\r*\N* \ifnum\r*<2\r*2\fi
\m**\r* \divide\m**2 \l**\r* \advance\l**-\m** \N*\n* \divide\N*\r*
\k**\r* \multiply\k**\N* \dn*\n* \advance\dn*-\k** \divide\dn*2\advance\dn*\*one
\r*\l** \divide\r*2\advance\dn*\r* \advance\N*-2\k**#2\relax
\ifnum\l**<6{\rotate(#2)\mov(#1,0){\sm*}}\advance\k**\dn*
{\rotate(\k**)\mov(#1,0){\sl*}}\advance\k**\m**
{\rotate(\k**)\mov(#1,0){\sm*}}\loop
\advance\k**\l**{\rotate(\k**)\mov(#1,0){\sl*}}\advance\k**\m**
{\rotate(\k**)\mov(#1,0){\sm*}}\advance\N*-1\ifnum\N*>0\repeat
{\rotate(#3)\mov(#1,0){\sl*}}\else\advance\k**\dn*
\arc(#1)[#2,\k**]\loop\advance\k**\m** \r*\k**
\advance\k**\l** {\arc(#1)[\r*,\k**]}\relax
\advance\N*-1\ifnum\N*>0\repeat
\advance\k**\m**\arc(#1)[\k**,#3]\fi}}

\def\triangarc#1(#2)[#3,#4]{{\k**=#3\n*=#4\advance\n*-\k**
\L*=1000sp\L*#2\L* \multiply\L*\n* \multiply\L*\Nhalfperiods
\divide\L*57\N*\L* \divide\N*1000\ifnum\N*=0\N*1\fi
\d**=#2\Lengthunit \d*\d** \divide\d*57\multiply\d*\n*
\r*\n*  \divide\r*\N* \ifnum\r*<2\r*2\fi
\m**\r* \divide\m**2 \l**\r* \advance\l**-\m** \N*\n* \divide\N*\r*
\dt*\d* \divide\dt*\N* \dt*.5\dt* \dt*#1\dt*
\divide\dt*1000\multiply\dt*\magnitude
\k**\r* \multiply\k**\N* \dn*\n* \advance\dn*-\k** \divide\dn*2\relax
\r*\l** \divide\r*2\advance\dn*\r* \advance\N*-1\k**#3\relax
{\rotate(#3)\mov(#2,0){\sm*}}\advance\k**\dn*
{\rotate(\k**)\mov(#2,0){\sl*}}\advance\k**-\m**\advance\l**\m**\loop\dt*-\dt*
\d*\d** \advance\d*\dt*
\advance\k**\l**{\rotate(\k**)\rmov*(\d*,0pt){\sl*}}%
\advance\N*-1\ifnum\N*>0\repeat\advance\k**\m**
{\rotate(\k**)\mov(#2,0){\sl*}}{\rotate(#4)\mov(#2,0){\sl*}}}}

\def\wavearc#1(#2)[#3,#4]{{\k**=#3\n*=#4\advance\n*-\k**
\L*=4000sp\L*#2\L* \multiply\L*\n* \multiply\L*\Nhalfperiods
\divide\L*57\N*\L* \divide\N*1000\ifnum\N*=0\N*1\fi
\d**=#2\Lengthunit \d*\d** \divide\d*57\multiply\d*\n*
\r*\n*  \divide\r*\N* \ifnum\r*=0\r*1\fi
\m**\r* \divide\m**2 \l**\r* \advance\l**-\m** \N*\n* \divide\N*\r*
\dt*\d* \divide\dt*\N* \dt*.7\dt* \dt*#1\dt*
\divide\dt*1000\multiply\dt*\magnitude
\k**\r* \multiply\k**\N* \dn*\n* \advance\dn*-\k** \divide\dn*2\relax
\divide\N*4\advance\N*-1\k**#3\relax
{\rotate(#3)\mov(#2,0){\sm*}}\advance\k**\dn*
{\rotate(\k**)\mov(#2,0){\sl*}}\advance\k**-\m**\advance\l**\m**\loop\dt*-\dt*
\d*\d** \advance\d*\dt* \dd*\d** \advance\dd*1.3\dt*
\advance\k**\r*{\rotate(\k**)\rmov*(\d*,0pt){\sl*}}\relax
\advance\k**\r*{\rotate(\k**)\rmov*(\dd*,0pt){\sl*}}\relax
\advance\k**\r*{\rotate(\k**)\rmov*(\d*,0pt){\sl*}}\relax
\advance\k**\r*
\advance\N*-1\ifnum\N*>0\repeat\advance\k**\m**
{\rotate(\k**)\mov(#2,0){\sl*}}{\rotate(#4)\mov(#2,0){\sl*}}}}

\def\gmov*#1(#2,#3)#4{\rlap{\L*=#1\Lengthunit
\xL*=#2\L* \yL*=#3\L*
\rx* \gcos*\xL* \tmp* \gsin*\yL* \advance\rx*-\tmp*
\ry* \gcos*\yL* \tmp* \gsin*\xL* \advance\ry*\tmp*
\rx*=\xscale\rx* \ry*=\yscale\ry*
\xL* \the\cos*\rx* \tmp* \the\sin*\ry* \advance\xL*-\tmp*
\yL* \the\cos*\ry* \tmp* \the\sin*\rx* \advance\yL*\tmp*
\kern\xL*\raise\yL*\hbox{#4}}}

\def\rgmov*(#1,#2)#3{\rlap{\xL*#1\yL*#2\relax
\rx* \gcos*\xL* \tmp* \gsin*\yL* \advance\rx*-\tmp*
\ry* \gcos*\yL* \tmp* \gsin*\xL* \advance\ry*\tmp*
\rx*=\xscale\rx* \ry*=\yscale\ry*
\xL* \the\cos*\rx* \tmp* \the\sin*\ry* \advance\xL*-\tmp*
\yL* \the\cos*\ry* \tmp* \the\sin*\rx* \advance\yL*\tmp*
\kern\xL*\raise\yL*\hbox{#3}}}

\def\Earc(#1)[#2,#3][#4,#5]{{\k*=#2\l*=#3\m*=\l*
\advance\m*-6\ifnum\k*>\l*\relax\else\def\xscale{#4}\def\yscale{#5}\relax
{\angle**0\rotate(#2)}\gmov*(#1,0){\sm*}\loop
\ifnum\k*<\m*\advance\k*5\relax
{\angle**0\rotate(\k*)}\gmov*(#1,0){\sl*}\repeat
{\angle**0\rotate(#3)}\gmov*(#1,0){\sl*}\relax
\def\xscale{1}\def\yscale{1}\fi}}

\def\dashEarc(#1)[#2,#3][#4,#5]{{\k**=#2\n*=#3\advance\n*-1\advance\n*-\k**
\L*=1000sp\L*#1\L* \multiply\L*\n* \multiply\L*\Nhalfperiods
\divide\L*57\N*\L* \divide\N*2000\ifnum\N*=0\N*1\fi
\r*\n*  \divide\r*\N* \ifnum\r*<2\r*2\fi
\m**\r* \divide\m**2 \l**\r* \advance\l**-\m** \N*\n* \divide\N*\r*
\k**\r*\multiply\k**\N* \dn*\n* \advance\dn*-\k** \divide\dn*2\advance\dn*\*one
\r*\l** \divide\r*2\advance\dn*\r* \advance\N*-2\k**#2\relax
\ifnum\l**<6\def\xscale{#4}\def\yscale{#5}\relax
{\angle**0\rotate(#2)}\gmov*(#1,0){\sm*}\advance\k**\dn*
{\angle**0\rotate(\k**)}\gmov*(#1,0){\sl*}\advance\k**\m**
{\angle**0\rotate(\k**)}\gmov*(#1,0){\sm*}\loop
\advance\k**\l**{\angle**0\rotate(\k**)}\gmov*(#1,0){\sl*}\advance\k**\m**
{\angle**0\rotate(\k**)}\gmov*(#1,0){\sm*}\advance\N*-1\ifnum\N*>0\repeat
{\angle**0\rotate(#3)}\gmov*(#1,0){\sl*}\def\xscale{1}\def\yscale{1}\else
\advance\k**\dn* \Earc(#1)[#2,\k**][#4,#5]\loop\advance\k**\m** \r*\k**
\advance\k**\l** {\Earc(#1)[\r*,\k**][#4,#5]}\relax
\advance\N*-1\ifnum\N*>0\repeat
\advance\k**\m**\Earc(#1)[\k**,#3][#4,#5]\fi}}

\def\triangEarc#1(#2)[#3,#4][#5,#6]{{\k**=#3\n*=#4\advance\n*-\k**
\L*=1000sp\L*#2\L* \multiply\L*\n* \multiply\L*\Nhalfperiods
\divide\L*57\N*\L* \divide\N*1000\ifnum\N*=0\N*1\fi
\d**=#2\Lengthunit \d*\d** \divide\d*57\multiply\d*\n*
\r*\n*  \divide\r*\N* \ifnum\r*<2\r*2\fi
\m**\r* \divide\m**2 \l**\r* \advance\l**-\m** \N*\n* \divide\N*\r*
\dt*\d* \divide\dt*\N* \dt*.5\dt* \dt*#1\dt*
\divide\dt*1000\multiply\dt*\magnitude
\k**\r* \multiply\k**\N* \dn*\n* \advance\dn*-\k** \divide\dn*2\relax
\r*\l** \divide\r*2\advance\dn*\r* \advance\N*-1\k**#3\relax
\def\xscale{#5}\def\yscale{#6}\relax
{\angle**0\rotate(#3)}\gmov*(#2,0){\sm*}\advance\k**\dn*
{\angle**0\rotate(\k**)}\gmov*(#2,0){\sl*}\advance\k**-\m**
\advance\l**\m**\loop\dt*-\dt* \d*\d** \advance\d*\dt*
\advance\k**\l**{\angle**0\rotate(\k**)}\rgmov*(\d*,0pt){\sl*}\relax
\advance\N*-1\ifnum\N*>0\repeat\advance\k**\m**
{\angle**0\rotate(\k**)}\gmov*(#2,0){\sl*}\relax
{\angle**0\rotate(#4)}\gmov*(#2,0){\sl*}\def\xscale{1}\def\yscale{1}}}

\def\waveEarc#1(#2)[#3,#4][#5,#6]{{\k**=#3\n*=#4\advance\n*-\k**
\L*=4000sp\L*#2\L* \multiply\L*\n* \multiply\L*\Nhalfperiods
\divide\L*57\N*\L* \divide\N*1000\ifnum\N*=0\N*1\fi
\d**=#2\Lengthunit \d*\d** \divide\d*57\multiply\d*\n*
\r*\n*  \divide\r*\N* \ifnum\r*=0\r*1\fi
\m**\r* \divide\m**2 \l**\r* \advance\l**-\m** \N*\n* \divide\N*\r*
\dt*\d* \divide\dt*\N* \dt*.7\dt* \dt*#1\dt*
\divide\dt*1000\multiply\dt*\magnitude
\k**\r* \multiply\k**\N* \dn*\n* \advance\dn*-\k** \divide\dn*2\relax
\divide\N*4\advance\N*-1\k**#3\def\xscale{#5}\def\yscale{#6}\relax
{\angle**0\rotate(#3)}\gmov*(#2,0){\sm*}\advance\k**\dn*
{\angle**0\rotate(\k**)}\gmov*(#2,0){\sl*}\advance\k**-\m**
\advance\l**\m**\loop\dt*-\dt*
\d*\d** \advance\d*\dt* \dd*\d** \advance\dd*1.3\dt*
\advance\k**\r*{\angle**0\rotate(\k**)}\rgmov*(\d*,0pt){\sl*}\relax
\advance\k**\r*{\angle**0\rotate(\k**)}\rgmov*(\dd*,0pt){\sl*}\relax
\advance\k**\r*{\angle**0\rotate(\k**)}\rgmov*(\d*,0pt){\sl*}\relax
\advance\k**\r*
\advance\N*-1\ifnum\N*>0\repeat\advance\k**\m**
{\angle**0\rotate(\k**)}\gmov*(#2,0){\sl*}\relax
{\angle**0\rotate(#4)}\gmov*(#2,0){\sl*}\def\xscale{1}\def\yscale{1}}}

\newcount\CatcodeOfAtSign
\CatcodeOfAtSign=\the\catcode`\@
\catcode`\@=11
\def\@arc#1[#2][#3]{\rlap{\Lengthunit=#1\Lengthunit
\sm*\l*arc(#2.1914,#3.0381)[#2][#3]\relax
\mov(#2.1914,#3.0381){\l*arc(#2.1622,#3.1084)[#2][#3]}\relax
\mov(#2.3536,#3.1465){\l*arc(#2.1084,#3.1622)[#2][#3]}\relax
\mov(#2.4619,#3.3086){\l*arc(#2.0381,#3.1914)[#2][#3]}}}

\def\dash@arc#1[#2][#3]{\rlap{\Lengthunit=#1\Lengthunit
\d*arc(#2.1914,#3.0381)[#2][#3]\relax
\mov(#2.1914,#3.0381){\d*arc(#2.1622,#3.1084)[#2][#3]}\relax
\mov(#2.3536,#3.1465){\d*arc(#2.1084,#3.1622)[#2][#3]}\relax
\mov(#2.4619,#3.3086){\d*arc(#2.0381,#3.1914)[#2][#3]}}}

\def\wave@arc#1[#2][#3]{\rlap{\Lengthunit=#1\Lengthunit
\w*lin(#2.1914,#3.0381)\relax
\mov(#2.1914,#3.0381){\w*lin(#2.1622,#3.1084)}\relax
\mov(#2.3536,#3.1465){\w*lin(#2.1084,#3.1622)}\relax
\mov(#2.4619,#3.3086){\w*lin(#2.0381,#3.1914)}}}

\def\bezier#1(#2,#3)(#4,#5)(#6,#7){\N*#1\l*\N* \advance\l*\*one
\d* #4\Lengthunit \advance\d* -#2\Lengthunit \multiply\d* \*two
\b* #6\Lengthunit \advance\b* -#2\Lengthunit
\advance\b*-\d* \divide\b*\N*
\d** #5\Lengthunit \advance\d** -#3\Lengthunit \multiply\d** \*two
\b** #7\Lengthunit \advance\b** -#3\Lengthunit
\advance\b** -\d** \divide\b**\N*
\mov(#2,#3){\sm*{\loop\ifnum\m*<\l*
\a*\m*\b* \advance\a*\d* \divide\a*\N* \multiply\a*\m*
\a**\m*\b** \advance\a**\d** \divide\a**\N* \multiply\a**\m*
\rmov*(\a*,\a**){\unhcopy\spl*}\advance\m*\*one\repeat}}}

\catcode`\*=12

\newcount\n@ast
\def\n@ast@#1{\n@ast0\relax\get@ast@#1\end}
\def\get@ast@#1{\ifx#1\end\let\next\relax\else
\ifx#1*\advance\n@ast1\fi\let\next\get@ast@\fi\next}

\newif\if@up \newif\if@dwn
\def\up@down@#1{\@upfalse\@dwnfalse
\if#1u\@uptrue\fi\if#1U\@uptrue\fi\if#1+\@uptrue\fi
\if#1d\@dwntrue\fi\if#1D\@dwntrue\fi\if#1-\@dwntrue\fi}

\def\halfcirc#1(#2)[#3]{{\Lengthunit=#2\Lengthunit\up@down@{#3}\relax
\if@up\mov(0,.5){\@arc[-][-]\@arc[+][-]}\fi
\if@dwn\mov(0,-.5){\@arc[-][+]\@arc[+][+]}\fi
\def\lft{\mov(0,.5){\@arc[-][-]}\mov(0,-.5){\@arc[-][+]}}\relax
\def\rght{\mov(0,.5){\@arc[+][-]}\mov(0,-.5){\@arc[+][+]}}\relax
\if#3l\lft\fi\if#3L\lft\fi\if#3r\rght\fi\if#3R\rght\fi
\n@ast@{#1}\relax
\ifnum\n@ast>0\if@up\shade[+]\fi\if@dwn\shade[-]\fi\fi
\ifnum\n@ast>1\if@up\dshade[+]\fi\if@dwn\dshade[-]\fi\fi}}

\def\halfdashcirc(#1)[#2]{{\Lengthunit=#1\Lengthunit\up@down@{#2}\relax
\if@up\mov(0,.5){\dash@arc[-][-]\dash@arc[+][-]}\fi
\if@dwn\mov(0,-.5){\dash@arc[-][+]\dash@arc[+][+]}\fi
\def\lft{\mov(0,.5){\dash@arc[-][-]}\mov(0,-.5){\dash@arc[-][+]}}\relax
\def\rght{\mov(0,.5){\dash@arc[+][-]}\mov(0,-.5){\dash@arc[+][+]}}\relax
\if#2l\lft\fi\if#2L\lft\fi\if#2r\rght\fi\if#2R\rght\fi}}

\def\halfwavecirc(#1)[#2]{{\Lengthunit=#1\Lengthunit\up@down@{#2}\relax
\if@up\mov(0,.5){\wave@arc[-][-]\wave@arc[+][-]}\fi
\if@dwn\mov(0,-.5){\wave@arc[-][+]\wave@arc[+][+]}\fi
\def\lft{\mov(0,.5){\wave@arc[-][-]}\mov(0,-.5){\wave@arc[-][+]}}\relax
\def\rght{\mov(0,.5){\wave@arc[+][-]}\mov(0,-.5){\wave@arc[+][+]}}\relax
\if#2l\lft\fi\if#2L\lft\fi\if#2r\rght\fi\if#2R\rght\fi}}

\catcode`\*=11

\def\Circle#1(#2){\halfcirc#1(#2)[u]\halfcirc#1(#2)[d]\n@ast@{#1}\relax
\ifnum\n@ast>0\L*=\xscale\Lengthunit
\ifnum\angle**=0\clap{\vrule width#2\L* height.1pt}\else
\L*=#2\L*\L*=.5\L*\special{em:linewidth .001pt}\relax
\rmov*(-\L*,0pt){\sm*}\rmov*(\L*,0pt){\sl*}\relax
\special{em:linewidth \the\linwid*}\fi\fi}

\catcode`\*=12

\def\wavecirc(#1){\halfwavecirc(#1)[u]\halfwavecirc(#1)[d]}

\def\dashcirc(#1){\halfdashcirc(#1)[u]\halfdashcirc(#1)[d]}

\def\xscale{1}
\def\yscale{1}

\def\Ellipse#1(#2)[#3,#4]{\def\xscale{#3}\def\yscale{#4}\relax
\Circle#1(#2)\def\xscale{1}\def\yscale{1}}

\def\dashEllipse(#1)[#2,#3]{\def\xscale{#2}\def\yscale{#3}\relax
\dashcirc(#1)\def\xscale{1}\def\yscale{1}}

\def\waveEllipse(#1)[#2,#3]{\def\xscale{#2}\def\yscale{#3}\relax
\wavecirc(#1)\def\xscale{1}\def\yscale{1}}

\def\halfEllipse#1(#2)[#3][#4,#5]{\def\xscale{#4}\def\yscale{#5}\relax
\halfcirc#1(#2)[#3]\def\xscale{1}\def\yscale{1}}

\def\halfdashEllipse(#1)[#2][#3,#4]{\def\xscale{#3}\def\yscale{#4}\relax
\halfdashcirc(#1)[#2]\def\xscale{1}\def\yscale{1}}

\def\halfwaveEllipse(#1)[#2][#3,#4]{\def\xscale{#3}\def\yscale{#4}\relax
\halfwavecirc(#1)[#2]\def\xscale{1}\def\yscale{1}}

\catcode`\@=\the\CatcodeOfAtSign

\section{Introduction}

The investigation of the energy levels of hydrogenic atoms with high
accuracy is important for the check of the Standard Model and extraction
the values of fundamental physical constants (the fine structure constant,
the electron and muon masses, the proton charge radius etc.) \cite{MT}.
The Zemach radius takes substantial role among many fundamental parameters
determining the structure of the energy levels of simple atomic systems
\cite{Zemach}. It is just this parameter which is responsible for the
hyperfine splitting (HFS) interval in electronic and muonic hydrogen.
The ground state hydrogen hyperfine splitting measurement was made many
years ago with very high accuracy \cite{Hellwig}:
\begin{equation}
\Delta \nu_{exp}^{HFS}(e p)=1~420~405.751~766~7(9)~kHz.
\end{equation}
The theoretical expression for the hyperfine splitting in the hydrogen
\begin{equation}
\Delta E_{theor}^{HFS}=E^F\left(1+\delta^{QED}+\delta^{str}+\delta^{pol}+
\delta^{HVP}\right),~~E^F=\frac{8}{3}\alpha^4\frac{\mu_Pm_1^2m_2^2}
{(m_1+m_2)^3},
\end{equation}
includes several corrections to the Fermi energy $E^F$:
$\delta^{QED}$ represents the QED contribution,
the corrections $\delta^{str}$ and $\delta^{pol}$ are the proton structure
and polarizability contributions,
$\delta^{HVP}$ is the contribution of hadronic vacuum polarization (HVP).
$\mu_p$ is the proton magnetic moment in nuclear magnetons, $m_1$ is the lepton
mass, $m_2$ is the proton mass.
The expression (2) is valid both for the muonic and electronic hydrogen
but the exact value of the corrections is essentially different for these
atoms. The proton structure and polarizability effects lead to main
theoretical uncertainty in the expression (2).
The investigation of the nuclear structure and polarizability corrections
in the hyperfine splitting of the energy levels was done in many papers
\cite{Zemach,Guerin,ZSFC,MPK,SGK,Friar,Pineda,FM3,DS,BE,BY,FM,KP1999,CFM} (see complete
discussion in the review article \cite{EGS}). It is valid to say that during
several tens of years the theory fall behind the experiment in solving
this task. Existing difference between the theory and experiment without
accounting the proton polarizability contribution can be expressed as follows
\cite{EGS}:
\begin{equation}
\frac{\Delta E^{HFS}_{theor}(e~p)-\Delta E^{exp}_{HFS}(e~p)}{E^F(e~p)}=
-4.5(1.1)\times 10^{-6}.
\end{equation}
There exists at least two possibilities to improve theoretical expression (2).
First possibility is connected with new more exact measurement of the electric
$G_E$ and magnetic $G_M$ proton form factors. The integral over the product of
these form factors gives main part of the proton structure correction
$\delta^{str}$ (the Zemach correction \cite{Zemach}):
\begin{equation}
\Delta E_Z=E^F\frac{2\mu\alpha}{\pi^2}\int \frac{d{\bf p}}{{\bf p}^4}
\Biggl[\frac{G_E(-{\bf p}^2)G_M(-{\bf p}^2)}{\mu_P}-1\Biggr]=E^F(-2\mu\alpha)
R_p,~W=\alpha\mu,
\end{equation}
where $\mu$ is the reduced mass of two particles, $R_p$ is the Zemach radius.
In the coordinate representation the Zemach correction (4) is determined by
the contraction of the charge $\rho_E$ and magnetic moment $\rho_M$
distributions. The Zemach radius represents the integral characteristic of
the proton structure effects in the hyperfine splitting of the energy levels.
It may be considered as new fundamental proton parameter in the hydrogen atom.

Another possibility consists in the using the muonic hydrogen. The measurement
of the hyperfine splitting of the $1S$ and $2S$ energy levels with the accuracy
$10^{-5}$ in the muonic hydrogen \cite{Bakalov,BMR,Dupays} allows to obtain
the value of the Zemach radius with the accuracy $10^{-3}$. Then it can be
used for obtaining new theoretical value for the hyperfine splitting in the
electronic hydrogen and new theoretical restriction on the value of the proton
polarizability contribution. An attempts to obtain the value of the Zemach
radius were made recently using just as the experimental data on elastic
electron-proton scattering so the hyperfine splitting in
the hydrogen atom (see the Table 1 and the references \cite{Dupays,Volotka,
FS,Brodsky}). The difference between corresponding
results reaches $7\%$.

\begin{table}
\caption{\label{t1} The values of the Zemach radii.}
\bigskip
\begin{ruledtabular}
\begin{tabular}{|c|c|c|c|c|}   \hline
The Zemach & Hyperfine splitting &Hyperfine splitting & Elastic e-p& Elastic e-p \\
radius    &in electronic hydrogen \cite{Dupays} &in electronic hydrogen \cite{Volotka} & scattering \cite{FS}& scattering \cite{Brodsky}      \\    \hline
$R_p$   & 1.047(19) fm & 1.040(16) fm & 1.086(12) fm & 1.013(16) fm \\ \hline
\end{tabular}
\end{ruledtabular}
\end{table}

The investigation of different contributions to the energy levels
of the muonic atoms was done many years ago in Ref. \cite{MS,BR,KP}.
So, at present there is need for new more complete analysis of all
possible corrections in the HFS of the $\mu p$  with the declared
accuracy $10^{-5}$. The corrections of order $\alpha^5$ to the
hyperfine splitting of the $2S$ state and the Lamb shift $2P-2S$
in the $\mu p$ were calculated in Refs.\cite{KP,VP2004}. In this
study we continue the investigation of the different contributions of orders
$\alpha^5$ and $\alpha^6$ to the muonic hydrogen HFS of the $S$ states
which was begun in Ref.\cite{FM2004}.
The aim of the work consists in obtaining the numerical value of the $2S$ HFS
with the accuracy $10^{-5}$ and the Sternheim's hyperfine splitting interval
$[8\Delta E^{HFS}(2S)-\Delta E^{HFS}(1S)]$ \cite{MS0}
in the muonic hydrogen with the accuracy $10^{-6}$ which
can serve as reliable guide for corresponding experiment and the measurement
$2P-2S$ Lamb shift \cite{Kottmann}.
Basic problems of the HFS measurement in the muonic hydrogen were
discussed in Refs.\cite{Bakalov,Dupays}.

\section{Effects of vacuum polarization in the one-photon interaction}

Our calculation of different energy levels of the hydrogen-like atoms are
carried out on the basis of the quasipotential approach where the
two-particle bound state is described by the Schroedinger-type equation
\cite{MF85}:

\begin{eqnarray}
\left[G^f\right]^{-1}\psi_M\equiv\left(\frac{b^2}{2\mu_R}-\frac{{\bf p}^2}{2\mu_R}\right)
\psi_M({\bf p})=
\int\frac{d{\bf q}}{(2\pi)^3}V({\bf p},{\bf q},M)\psi_M({\bf q}),
\end{eqnarray}
where
\begin{displaymath}
b^2=E_1^2-m_1^2=E_2^2-m_2^2,
\end{displaymath}
$\mu_R=E_1E_2/M$ is the relativistic reduced mass, $M=E_1+E_2$ is the bound state mass.
The quasipotential of the equation (5) is constructed in the quantum electrodynamics
by the perturbative
series using projected on positive states the two-particle off mass shell
scattering amplitude $T$ at zero relative energies of the particles:

\begin{equation}
V=V^{(1)}+V^{(2)}+V^{(3)}+...,~~~~~T=T^{(1)}+T^{(2)}+T^{(3)}+...,
\end{equation}

\begin{equation}
V^{(1)}=T^{(1)},~V^{(2)}=T^{(2)}-T^{(1)}\times G^f\times T^{(1)}, ...~~.
\end{equation}

\begin{figure}
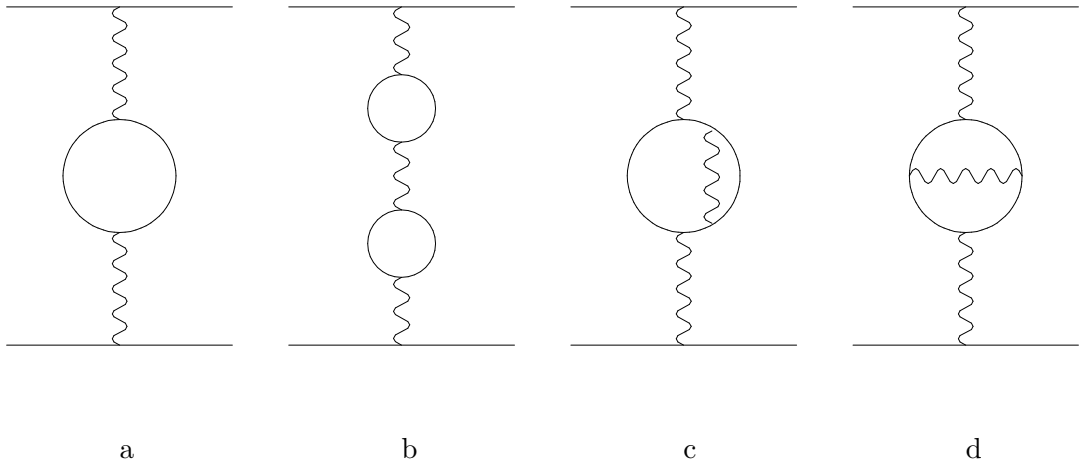

\magnitude=2000
\GRAPH(hsize=15){
\mov(0,0){\lin(2,0)}%
\mov(0,3){\lin(2,0)}%
\mov(2.5,0){\lin(2,0)}%
\mov(2.5,3){\lin(2,0)}%
\mov(5,0){\lin(2,0)}%
\mov(5,3){\lin(2,0)}%
\mov(7.5,0){\lin(2,0)}%
\mov(7.5,3){\lin(2,0)}%
\mov(1,1.5){\Circle(1.)}%
\mov(1,0){\wavelin(0,1)}%
\mov(1,2){\wavelin(0,1)}%
\mov(3.5,0){\wavelin(0,0.6)}%
\mov(3.5,1.2){\wavelin(0,0.6)}%
\mov(3.5,2.4){\wavelin(0,0.6)}%
\mov(3.5,0.9){\Circle(0.6)}%
\mov(3.5,2.1){\Circle(0.6)}%
\mov(6,0){\wavelin(0,1)}%
\mov(6,2){\wavelin(0,1)}%
\mov(6,1.5){\Circle(1.)}%
\mov(6.25,1.08){\wavelin(0,0.82)}%
\mov(8.5,0){\wavelin(0,1.)}%
\mov(8.5,2.){\wavelin(0,1)}%
\mov(8.,1.5){\wavelin(1,0)}%
\mov(8.5,1.5){\Circle(1.0)}%
\mov(1.,-1.){a}%
\mov(3.5,-1.){b}%
\mov(6.,-1.){c}%
\mov(8.5,-1.){d}%
}
\vspace{3mm}
\caption{Effects of the one- and two-loop vacuum polarization in the
one-photon interaction.}
\end{figure}

We take the ordinary Coulomb potential as initial approximation for the quasipotential
$V({\bf p},{\bf q},M)$: $V({\bf p},{\bf q},M)=V^C({\bf p}-{\bf q})+
\Delta V({\bf p},{\bf q},M).$

The increase of the lepton mass when we change the electronic hydrogen to the
muonic hydrogen
leads to the decrease of the Bohr radius in the $\mu p$. As a result the electron Compton wave length
and the Bohr radius are of the same order:
\begin{equation}
\frac{\hbar^2}{\mu e^2}:\frac{\hbar}{m_ec}=0.737384
\end{equation}
($m_e$ is the electron mass, $\mu$ is the reduced mass in the atom $\mu p$).
An important consequence of last relation is the increase the role of the
electron vacuum polarization effects in the energy spectrum of the $\mu p$
\cite{t4}. The effects of the vacuum polarization in the one-photon
interaction are shown in Fig.1.

To obtain the contribution of the diagram (a) Fig.1 (the electron vacuum polarization) to the
interaction operator there is need to make the following substitution in the photon
propagator \cite{t4}:

\begin{equation}
\frac{1}{k^2}\to \frac{\alpha}{\pi}\int_0^1 dv\frac{v^2\left(1-
\frac{v^2}{3}\right)}{k^2(1-v^2)-4m_e^2}.
\end{equation}
At $(-k^2)={\bf k}^2\sim\mu_e^2(Z\alpha)^2\sim m_e^2(Z\alpha)^2$ (electronic
hydrogen, $\mu_e$ is the reduced mass in hydrogen atom) we obtain
$-\alpha/15\pi m_e^2$ omitting the first term in the denominator of right
part of Eq.(9). But when ${\bf k}^2\sim \mu^2(Z\alpha)^2\sim m_1^2(Z\alpha)^2$
(muonic hydrogen, $m_1$ is the muon mass) than $\mu \alpha$ and $m_e$
are of the same order and it is impossible to use expansion over $\alpha$ in
the denominator of Eq.(9). To construct the hyperfine part of the
quasipotential in this case (the muonic hydrogen) in the one-photon
interaction we must use exact expression (9). We take into account that the
appearance of the electron mass $m_e$ in the denominator of the amplitude
leads effectively to the decrease the order of the correction. It is well
known that the hyperfine splitting quasipotential has the form \cite{RNF}:

\begin{equation}
V_{1\gamma}^{HFS}({\bf k})=\frac{4\pi Z\alpha}{m_1m_2}\frac{1+\kappa}{4}
\frac{1}{{\bf k}^2}[(\mathstrut\bm{\sigma}_1
\mathstrut\bm{\sigma}_2){\bf k}^2-(\mathstrut\bm{\sigma}_1
{\bf k})(\mathstrut\bm{\sigma}_2{\bf k})].
\end{equation}
For the $S$-states
\begin{equation}
V_{1\gamma}^{HFS}({\bf k})=\frac{8\pi Z\alpha}{3m_1m_2}\frac{
\mathstrut\bm{\sigma}_1
\mathstrut\bm{\sigma}_2}{4}(1+\kappa),
\end{equation}
$\kappa$=1.792847337(29) is the proton anomalous magnetic moment. Averaging
the potential (11) over the
Coulomb wave functions we obtain main contribution of order $(Z\alpha)^4$
to the HFS of the $2S$-state in the system $\mu p$ (the Fermi energy):
\begin{equation}
E^F(2S)=\frac{1}{3}(Z\alpha)^4\frac{\mu^3}{m_1m_2}(1+\kappa)=22.8054~meV.
\end{equation}

The modification of the Coulomb potential due to the vacuum polarization (VP)
is determined by means of Eq.(9) in the momentum representation
as follows \cite{t4}:
\begin{equation}
V^C_{VP}({\bf k})=-4\pi Z\alpha\frac{\alpha}{\pi}\int_1^\infty\frac{\sqrt
{\xi^2-1}}{3\xi^4}\frac{(2\xi^2+1)}{{\bf k}^2+4m_e^2\xi^2}d\xi
\end{equation}
In the coordinate representation we obtain:
\begin{equation}
V^C_{VP}(r)=\frac{\alpha}{3\pi}\int_1^\infty d\xi\frac{\sqrt{\xi^2-1}(
2\xi^2+1)}{\xi^4}\left(-\frac{Z\alpha}{r}e^{-2m_e\xi r}\right).
\end{equation}
The contribution of the electron vacuum polarization to the hyperfine
splitting part of the $1\gamma$ quasipotential for the $S$-states can be
derived in a similar way in the momentum and coordinate representations:
\begin{equation}
V_{1\gamma,~VP}^{HFS}({\bf k})=\frac{8\pi Z\alpha(1+\kappa)}{3m_1m_2}
\frac{(\mathstrut\bm{\sigma}_1\mathstrut\bm{\sigma}_2)}{4}{\bf k}^2\frac
{\alpha}{\pi}\int_1^\infty\frac{\sqrt{\xi^2-1}(2\xi^2+1)}{3\xi^4({\bf k}^2+4
m_e^2\xi^2)}d\xi,
\end{equation}
\begin{equation}
V_{1\gamma,~VP}^{HFS}(r)=\frac{8Z\alpha(1+\kappa)}{3m_1m_2}\frac
{(\mathstrut\bm{\sigma}_1
\mathstrut\bm{\sigma}_2)}{4}\frac{\alpha}{\pi}\int_1^\infty\frac{\sqrt{\xi^2-
1}(2\xi^2+1)}{3\xi^4}d\xi\left[\pi\delta({\bf r})-\frac{m_e^2\xi^2}{r}e^{-2m_e\xi r}
\right].
\end{equation}

Using Eq.(16) we can obtain the electron vacuum polarization
correction of order $\alpha^5$ to the HFS in the $\mu p$. Taking
the wave function of the $2S$-state
\begin{equation}
\psi_{200}(r)=\frac{W^{3/2}}{2\sqrt{2\pi}}e^{-Wr/2}\left(1-\frac{Wr}{2}\right),
~~~W=\mu Z\alpha,
\end{equation}
we represent this correction in the form:
\begin{equation}
\Delta E_{1\gamma,VP}{HFS}=\frac{\mu^3(Z\alpha)^4(1+\kappa)}{3m_1m_2}
\frac{\alpha}{\pi}\frac{m_e^3}{3W^3}\int_{m_e/W}^\infty\frac{\sqrt{\frac{W^2}
{m_e^2}\xi^2-1}}{\xi^4}\left(2\frac{W^2}{m_e^2}\xi^2+1\right)d\xi\times
\end{equation}
\begin{displaymath}
\times\left[
1-\int_0^\infty e^{-r(2\xi+1)/2\xi}\left(1-\frac{r}{4}\right)^2r dr\right]=
0.0481~meV.
\end{displaymath}
The contribution of the muon vacuum polarization (MVP) can be found
by means (16) after the substitution $m_e\to m_1$. This correction is of
order $\alpha^6$ due to the reason mentioned above. Numerical value is equal
\begin{equation}
\Delta E_{1\gamma,~MVP}^{HFS}=E^F(2S)\frac{3}{8}\frac{\mu}{m_1}Z\alpha^2=0.0004~ meV.
\end{equation}
In this expression and below $E^F(2S)$ is the Fermi energy of the $2S$ state.
The diagrams of the two-loop electron vacuum polarization shown in Fig.1
(b,c,d) give the contributions of the same order $\alpha^6$. The interaction
operator corresponding to the loop after loop amplitude can be obtained
using the relation (9). In the coordinate representation
\begin{equation}
V_{1\gamma,~VP-VP}^{HFS}(r)=\frac{8\pi Z\alpha(1+\kappa)}{3m_1m_2}
\frac{(\mathstrut\bm{\sigma}_1
\mathstrut\bm{\sigma}_2)}{4}\left(\frac{\alpha}{\pi}\right)^2
\int_1^\infty\frac{\sqrt{\xi^2-1}(2\xi^2+1)}{3\xi^4}d\xi\times
\end{equation}
\begin{displaymath}
\times\int_1^\infty\frac{\sqrt{\eta^2-1}(2\eta^2+1)}{3\eta^4}d\eta\left[
\delta({\bf r})-\frac{m_e^2}{\pi r(\eta^2-\xi^2)}\left(\eta^4e^{-2m_e\eta r}-
\xi^4e^{-2m_e\xi r}\right)\right],
\end{displaymath}
and the contribution to the energy spectrum
\begin{equation}
\Delta E_{1\gamma,~VP-VP}^{HFS}=0.0002~ meV.
\end{equation}
To calculate the contributions of the diagrams b, c in Fig.1 which are
determined by the polarization operator of the second order it is necessary
to make the substitution in the photon propagator \cite{EGS1}:
\begin{equation}
\frac{1}{k^2}\to \left(\frac{\alpha}{\pi}\right)^2\int_0^1\frac{f(v)}
{4m_e^2+k^2(1-v^2)}dv=\left(\frac{\alpha}{\pi}\right)^2\frac{2}{3}\int_0^1 dv
\frac{v}{4m_e^2+k^2(1-v^2)}\times
\end{equation}
\begin{displaymath}
\times\Biggl\{(3-v^2)(1+v^2)\left[Li_2\left(-\frac{1-v}{1+v}\right)+2Li_2
\left(\frac{1-v}{1+v}\right)+\frac{3}{2}\ln\frac{1+v}{1-v}\ln\frac{1+v}{2}-
\ln\frac{1+v}{1-v}\ln v\right]+
\end{displaymath}
\begin{displaymath}
\left[\frac{11}{16}(3-v^2)(1+v^2)+\frac{v^4}{4}\right]\ln\frac{1+v}{1-v}+
\left[\frac{3}{2}v(3-v^2)\ln\frac{1-v^2}{4}-2v(3-v^2)\ln v\right]+
\frac{3}{8}v(5-3v^2)\Biggr\}.
\end{displaymath}

To find numerical value of this correction we write the quasipotential
in the coordinate space:
\begin{equation}
\Delta V_{1\gamma,~ 2-loop~VP}^{HFS})=\frac{8\pi
Z\alpha(1+\kappa)}
{3m_1m_2}\left(\frac{\alpha}{\pi}\right)^2\int_0^1\frac{f(v)dv}{(1-v^2)}
\left[\delta({\bf r})-\frac{m_e^2}{\pi r
(1-v^2)}e^{-\frac{2m_er}{\sqrt {1-v^2}}}\right].
\end{equation}

The potential (23) gives the contribution to the $2S$ HFS in the muonic
hydrogen
\begin{equation}
\Delta E_{1\gamma,~ 2-loop~VP}^{HFS}=0.0001~ meV.
\end{equation}
We calculate all contribution numerically and the results are presented
with the accuracy 0.0001 meV.

\section{Second order of the perturbation theory}

The corrections of the second order of the perturbative series in the energy
spectrum are defined by the reduced Coulomb Green function (RCGF) \cite{VP}:

\begin{equation}
\tilde G_2({\bf r}, {\bf r'})=\sum_{l,m}\tilde g_{nl}(r,r')Y_{lm}({\bf n})
Y_{lm}^\ast({\bf n'}).
\end{equation}
The radial wave function $\tilde g_{nl}(r,r')$ was obtained in
Ref.\cite{VP} as an expansion over the Laguerre polynomials. For
the $2S$ - state
\begin{equation}
\tilde g_{20}(r,r')=-2\mu^2
Z\alpha\left[e^{-\frac{x+x'}{2}}\sum_{m=1,m\not=2}^\infty
\frac{L_{m-1}^1(x)L_{m-1}^1(x')}{m(m-2)}+\left(\frac{5}{2}+x\frac{\partial}
{\partial x}+x'\frac{\partial}{\partial
x'}\right)e^{-\frac{x+x'}{2}} L_1^1(x)L_1^1(x')\right],
\end{equation}
where $x=\mu Z\alpha r$, $L_n^m$ are the Laguerre polynomials:
\begin{equation}
L_n^m(x)=\frac{e^xx^{-m}}{n!}\left(\frac{d}{dx}\right)^n\left(e^{-x}
x^{n+m}\right).
\end{equation}
Some terms of the quasipotential contain the $\delta({\bf r})$ so we have to
know the quantity $\tilde G_2({\bf r},0)$. The expression for the RCGF was
found in this case in Ref.\cite{KI} on the basis of the Hoestler
representation for the Coulomb Green function after the subtraction the pole
term:
\begin{equation}
\tilde G_{2S}({\bf r},0)=-\frac{Z\alpha\mu^2}{4\pi}\frac{e^{-x/2}}{2x}
\left[4x(x-2)(\ln x+C)+x^3-13x^2+6x+4\right],
\end{equation}
where $C=0.5772...$ is the Euler constant. The main contribution of order
$\alpha^5$ in the second order of the perturbation theory can be written
in general form:
\begin{equation}
\Delta
E_{1~SOPT}^{HFS}=\sum_{n=1,n\not=2}^\infty\frac{<\psi_2^c|V_{VP}^C|
\psi_n^c><\psi_n^c|\Delta V_{1\gamma}^{HFS}|\psi_2^c>}
{E_2^c-E_n^c},
\end{equation}
where $\Delta V_{1\gamma}^{HFS}\sim\delta({\bf r})$. Using the relations (14),
(28) we can present Eq.(29) as follows:
\begin{equation}
\Delta E_{1~SOPT}^{HFS}=E^F(2S)(1+a_\mu)\frac{\alpha}{3\pi}\int_1^\infty d\xi
\frac{\sqrt{\xi^2-1}(2\xi^2+1)}{\xi^4}\times
\end{equation}
\begin{displaymath}
\times\int_0^\infty dx e^{-x(1+\frac{2m_e\xi}{W})}\left[4x(x-2)(\ln x+C)+x^3-
13x^2+6x+4\right]=0.0746~ meV.
\end{displaymath}
The contribution of order $\alpha^6$ in the second order of the perturbative
series which is determined by the vacuum polarization can be derived from
Eq.(29) changing $\Delta V_{1\gamma}^{HFS}\to\Delta V_{1\gamma~VP}^{HFS}$.
Using exact expressions for the wave function $\psi_2^c({\bf r})$ (17)
and the RCGF (28) we write this correction
\begin{equation}
\Delta E_{2~SOPT}^{HFS}=E^F(2S)\frac{\alpha^2}{9\pi^2}
\int_1^\infty d\xi\frac{\sqrt{\xi^2-1}(1+2\xi^2)}{\xi^4}
\times
\end{equation}
\begin{displaymath}
\times\int_1^\infty d\eta\frac{\sqrt{\eta^2-1}(1+2\eta^2)}{\eta^4}
\left[F_1(\xi)+F_2(\xi,\eta)+F_3(\xi,\eta)\right],
\end{displaymath}
\begin{equation}
F_1(\xi)=\frac{1}{a_1^5}\left[-12+23a_1-8a_1^2-4a_1^3+4a_1^4+4a_1(3-4a_1+
2a_1^2)\ln a_1\right], a_1=\left(1+\frac{2m_e}{W}\xi\right),
\end{equation}
\begin{displaymath}
F_2(\xi,\eta)=-\frac{2m_e^2\xi^2}{W^2}\frac{1}{a_2^5b_2^5}\Bigl[-12b_2(3-4b_2+
2b_2^2)+6a_2^2(-4+11b_2-8b_2^2+2b_2^3)-4a_2^2(-12+39b_2-
\end{displaymath}
\begin{displaymath}
-34b_2^2+12b_2^3)
+3a_2(-12+51b_2-52b_2^2+22b_2^3)\Bigr],~a_2=a_1,
b_2=\left(1+\frac{2m_e}{W}\eta\right),
\end{displaymath}
\begin{displaymath}
F_3(\xi,\eta)=-\frac{16m_e^2\xi^2}{W^2}\frac{a_3}{\left(1+\frac{2m_e\xi}{W}\right)^3
\left(1+\frac{2m_e\eta}{W}\right)^3}\Bigl\{\frac{4m_e^2}{W^2}\xi\eta\left[
\frac{a_3}{1-a_3}-2\ln(1-a_3)\right]+
\end{displaymath}
\begin{displaymath}
+\frac{1}{2}\left(\frac{\xi}{\eta}+\frac{\eta}{\xi}\right)\left[\frac{4a_3}
{1-a_3}+\frac{a_3^2}{(1-a_3)^2}-2\ln(1-a_3)\right]+\frac{W^2}{16m_e^2\xi\eta}
\Bigl[\frac{19a_3}{1-a_3}+\frac{6a_3^2}{(1-a_3)^2}+
\end{displaymath}
\begin{displaymath}
+\frac{a_3^2(1+a_3)}{(1-a_3)^3}-2\ln(1-a_3)\Bigr]\Bigr\},~a_3=\frac{1}
{\left(1+\frac{W}{2m_e\xi}\right)\left(1+\frac{W}{2m_e\eta}\right)}.
\end{displaymath}
Numerical value of this contribution is equal
\begin{equation}
\Delta E_{2~SOPT}^{HFS}=0.0003~ meV.
\end{equation}
The second order of the perturbative series gives also other relativistic
corrections of order $(Z\alpha)^6$ including recoil effects which were
studied in Ref.\cite{BYG,KN,FM1}. Corresponding numerical data are in the
Table 1.

\section{Proton structure and vacuum polarization effects}

The proton structure corrections in the system $\mu p$ are relatively large
in the comparison with the electronic hydrogen.
In the HFS of the muonic hydrogen these corrections are defined in the
leading order by the one-loop diagrams in Fig.2.

To construct the quasipotential corresponding to these diagrams we write
the proton tensor:

\begin{figure}
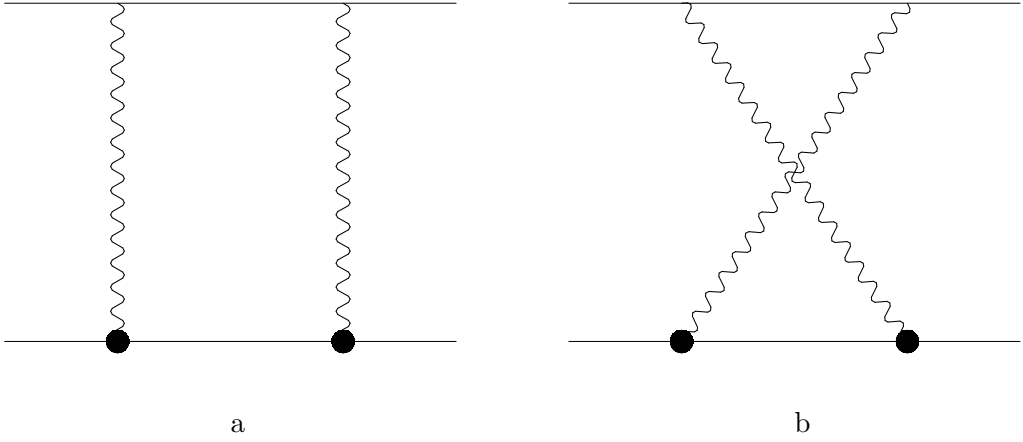

\magnitude=2000
\GRAPH(hsize=15){
\mov(0,0){\lin(1,0)}%
\mov(3,0){\lin(1,0)}%
\mov(0,3){\lin(4,0)}%
\mov(5,0){\lin(1,0)}%
\mov(8,0){\lin(1,0)}%
\mov(5,3){\lin(4,0)}%
\mov(3.,0.){\Circle**(0.2)}%
\mov(6.,0){\Circle**(0.2)}%
\mov(1.,0.){\Circle**(0.2)}%
\mov(8.,0.){\Circle**(0.2)}%
\mov(2.,-0.8){a}%
\mov(7.,-0.8){b}%
\mov(1.,0){\lin(2.,0)}%
\mov(6.,0){\lin(2.,0)}%
\mov(1,0){\wavelin(0,3)}%
\mov(3,0){\wavelin(0,3)}%
\mov(6,0){\wavelin(2,3)}%
\mov(8,0){\wavelin(-2,3)}%
}
\caption{Proton structure corrections of order $(Z\alpha)^5$. Bold circle
in the diagram represents the proton vertex operator.}
\end{figure}

\begin{equation}
M_{\mu\nu}^{(p)}=\bar u(q_2)\left[\gamma_\mu F_1+\frac{i}{2m_2}
\sigma_{\mu\omega}k^\omega F_2\right]\frac{\hat p_2-\hat k+m_2}{(p_2-k)^2-m_2^2+i0}
\left[\gamma_\nu F_1-\frac{i}{2m_2}\sigma_{\nu\lambda}k^\lambda
F_2\right]u(p_2),
\end{equation}
where $p_2, q_2$ are four momenta of the proton in initial and final states.
The construction of the potential can be essentially simplified using the
projection operators for the system muon-proton on the states with definite
spin:

\begin{equation}
\hat\pi(^1S_0)=[u(p_2)\bar v(p_1)]_{S=0}=\frac{(1+\gamma^0)}{2\sqrt{2}}
\gamma_5,~~~\hat\pi(^3S_1)=[u(p_2)\bar v(p_1)]_{S=1}=\frac{(1+\gamma^0)}
{2\sqrt{2}}\hat\epsilon.
\end{equation}
where $\epsilon^\mu$ is the polarization vector of the state with the spin 1.
Neglecting relative motion momenta of the particles in the initial and final
states we obtain
\begin{equation}
\Delta E_{str}^{HFS}=E^F(2S)\frac{Z\alpha m_1m_2}{8\pi (1+\kappa)}
\delta_{l0}\int\frac{id^4k}{\pi^2(k^2)^2}\Biggl[\frac{16k^6k_0^2}
{m_2^2}F_2^2+\frac{32k^8}{m_2^2}F_2^2-64k^2k_0^4F_2^2+
\end{equation}
\begin{displaymath}
+16k^4k_0^2F_1^2+128k^4k_0^2F_1F_2+64k^4k_0^2F_2^2+32k^6F_1^2+64k^6F_1F_2
\Biggr]\frac{1}{(k^4-4m_1^2k_0^2)(k^4-4m_2^2k_0^2)}.
\end{displaymath}
Transforming the integration in Eq.(36) to the Euclidean space
\begin{equation}
\int d^4k=4\pi\int k^3dk\int\sin^2\phi d\phi,~~k_0=k\cos\phi,
\end{equation}
we make analytical integration over the angle $\phi$ and present
the correction (36) as the one-dimensional integral over the variable k:
\begin{equation}
\Delta E_{str}^{HFS}=-E^F(2S)\frac{Z\alpha}{8\pi (1+\kappa)}\delta_{l0}
\int_0^\infty\frac{dk}{k}V(k),
\end{equation}
\begin{displaymath}
V(k)=\frac{2F_2^2k^2}{m_1m_2}+\frac{\mu}{(m_1-m_2)k(k+\sqrt{4m_1^2+k^2})}
\Biggl[-128F_1^2m_1^2-128F_1F_2m_1^2+16F_1^2k^2+
\end{displaymath}
\begin{displaymath}
+64F_1F_2k^2+16F_2^2k^2+\frac{32F_2^2m_1^2k^2}{m_2^2}+\frac{4F_2^2k^4}
{m_1^2}-\frac{4F_2^2k^4}{m_2^2}\Biggr]+\frac{\mu}{(m_1-m_2)k(k+
\sqrt{4m_2^2+k^2})}\times
\end{displaymath}
\begin{displaymath}
\times\left[128F_1^2m_2^2+128F_1F_2m_2^2-16F_1^2k^2-64F_1F_2k^2-48F_2^2k^2\right].
\end{displaymath}
To cancel infrared divergence in Eq.(38) it is necessary to add the
contribution of the iteration term of the quasipotential (10) in the HFS of
the $\mu p$:
\begin{equation}
\Delta E_{iter,str}^{HFS}=-<V_{1\gamma}\times G^f\times V_{1\gamma}>
_{str}^{HFS}=-\frac{8}{3}\frac{\mu^4(Z\alpha)^5(1+\kappa)}{m_1m_2\pi}
\int_0^\infty\frac{dk}{k^2},
\end{equation}
where the angular brackets represent the averaging of the interaction operator
over the Coulomb wave functions and the index HFS shows the hyperfine part in
the iteration term of the quasipotential (8). The sum of the expressions (38)
and (39) coincides with the result of Ref.\cite{KP}. The integration in
Eqs.(38) and (39) was done by means of the parameterization of the proton
electromagnetic form factors obtained from the analysis of elastic
lepton-nucleon scattering \cite{Simon}. Numerically the proton structure
correction of order $(Z\alpha)^5$ is equal
\begin{equation}
\Delta E_{str}^{HFS}+\Delta E_{iter,~str}^{HFS}=-0.1518~meV
\end{equation}
Moreover, the effects of the proton structure must be taken into account
carefully in the amplitudes of higher order over $\alpha$ shown in Fig.3.

\begin{figure}
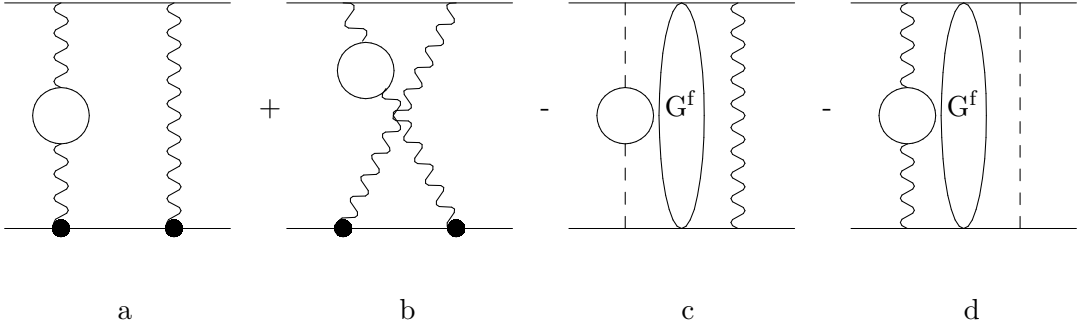

\magnitude=2000
\GRAPH(hsize=15){
\mov(0,0){\lin(2,0)}%
\mov(0,2){\lin(2,0)}%
\mov(2.5,0){\lin(2,0)}%
\mov(2.5,2){\lin(2,0)}%
\mov(5,0){\lin(2,0)}%
\mov(5,2){\lin(2,0)}%
\mov(7.5,0){\lin(2,0)}%
\mov(7.5,2){\lin(2,0)}%
\mov(0.5,0.){\Circle**(0.15)}%
\mov(1.5,0){\Circle**(0.15)}%
\mov(3.,0.){\Circle**(0.15)}%
\mov(4.,0.){\Circle**(0.15)}%
\mov(6,1){\Ellipse(0.4)[1,5]}%
\mov(5.85,1){${\rm G^f}$}%
\mov(8.5,1){\Ellipse(0.4)[1,5]}%
\mov(8.35,1){${\rm G^f}$}%
\mov(1.,-0.8){a}%
\mov(3.5,-0.8){b}%
\mov(6.,-0.8){c}%
\mov(8.5,-0.8){d}%
\mov(1.5,0){\wavelin(0,2)}%
\mov(3,0){\wavelin(1,2)}%
\mov(6.5,0){\wavelin(0,2)}%
\mov(9,0){\dashlin(0,2)}%
\mov(0.5,1.){\Circle(0.5)}%
\mov(0.5,0){\wavelin(0,0.75)}%
\mov(0.5,1.25){\wavelin(0,0.75)}%
\mov(2.25,1.){+}%
\mov(4.75,1.){-}%
\mov(7.25,1.){-}%
\mov(5.5,0){\dashlin(0,0.75)}%
\mov(5.5,1.25){\dashlin(0,0.75)}%
\mov(5.5,1.){\Circle(0.5)}%
\mov(8,0){\wavelin(0,0.75)}%
\mov(8,1.25){\wavelin(0,0.75)}%
\mov(8,1.){\Circle(0.5)}%
\mov(3,2){\wavelin(0.17,-0.34)}%
\mov(4,0){\wavelin(-0.62,1.24)}%
\mov(3.2,1.4){\Circle(0.5)}%
}
\caption{Vacuum polarization and proton structure corrections of order
$\alpha(Z\alpha)^5$. Dashed line in the diagram represents the Coulomb photon.}
\end{figure}

The contributions of the diagrams (a) and (b) in Fig.3 to the potential can
be found as for the amplitudes in Fig.2. taking into account the
transformation of one exchange photon propagator as in Eq.(9). Corresponding
correction to the HFS of the energy level is equal
\begin{equation}
\Delta E_{str,VP}^{HFS}=-E^F(2S)\frac{Z\alpha}{8\pi(1+\kappa)}2\frac{\alpha}
{\pi}\int_0^1\frac{v^2\left(1-\frac{v^2}{3}\right)dv}{k^2(1-v^2)+4m_e^2}\int_0
^\infty dk V_{VP}(k),
\end{equation}
where the potential $V_{VP}(k)$ differs from $V(k)$ in the relation (38)
only by the factor $k^2$. Despite of the finiteness of the integral (41)
the amplitude terms of the quasipotential in Fig.3 (a), (b) must be completed
by two iteration terms shown in Fig. 3 (c), (d). First addendum
$<V^c\times G^f\times\Delta V_{VP}^{HFS}>$ of order $\alpha(Z\alpha)^4$ must
be subtracted because the $2\gamma$ amplitudes (a) and (b) in Fig.3 produce
lower order contribution. Second term $<V^c_{VP}\times G^f\times
V_{1\gamma}^{HFS}>$ which is also of order $\alpha(Z\alpha)^4$ has the
structure similar to Eq.(29) of the second order of the perturbative series.
The contributions of discussed iteration terms to the HFS of the $\mu p$
coincide:
\begin{equation}
\Delta E_{iter,VP+str}^{HFS}=-2<V^c\times G^f\times\Delta V_{VP}^{HFS}>^{HFS}=
-2<V_{VP}^c\times G^f\times\Delta V_{1\gamma}^{HFS}>^{HFS}=
\end{equation}
\begin{displaymath}
=-E^F(2S)\frac{4(Z\alpha)
\mu\alpha}{m_e\pi^2}\int_0^\infty dk\int_0^1\frac{v^2\left(1-\frac{v^2}{3}\right)dv}
{k^2(1-v^2)+1}.
\end{displaymath}
Numerical value of the proton structure and vacuum polarization effects
of the $2\gamma$ amplitudes
\begin{equation}
\Delta E_{VP,str}^{HFS}+2\Delta E_{iter,VP+str}^{HFS}=-0.0026~meV.
\end{equation}
Hadronic vacuum polarization contribution to the HFS in the $\mu p$ was
studied in Ref.\cite{FM2}. Here we present it in the different form using the
expressions (38) and (41):
\begin{equation}
\Delta E_{HVP}^{HFS}=-E^F(2S)\frac{\alpha(Z\alpha)}{4\pi^2(1+\kappa)}
\int_{4m_\pi^2}^\infty\frac{\rho(s)ds}{k^2+s}\int_0^\infty dk V_{VP}(k).
\end{equation}
Dividing the integration range over $s$ on the intervals where the cross section
of the $e^+e^-$ annihilation into hadrons ($\rho(s)$ = $\sigma^h(e^+e^-
\to hadrons)/3s\sigma_{\mu\mu}$) is known from the experiment \cite{CMD} we
can make numerical integration in Eq.(44). Numerical value coincides with the
result obtained in Ref.\cite{FM2}:
\begin{equation}
\Delta E_{HVP}^{HFS}=0.0005~meV.
\end{equation}

\section{Proton structure effects, self energy and vertex corrections of
order $\alpha(Z\alpha)^5$}

\begin{figure}
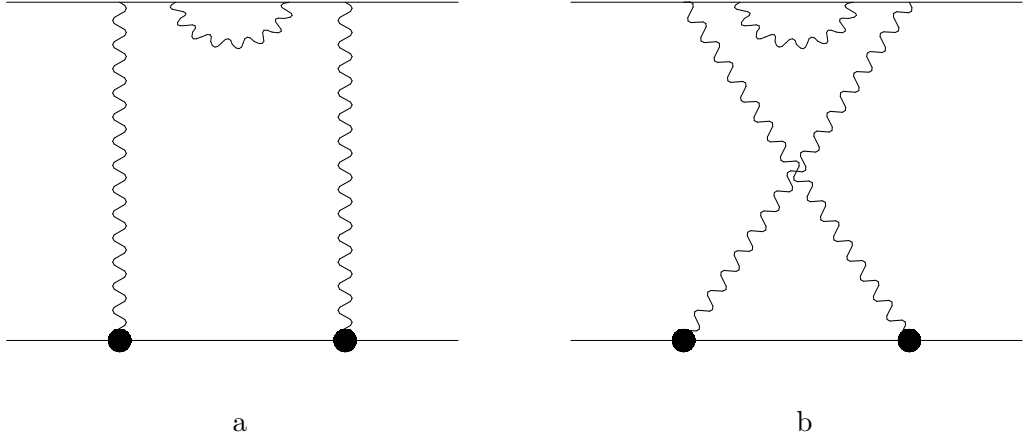

\magnitude=2000
\GRAPH(hsize=15){
\mov(0,0){\lin(1,0)}%
\mov(2,3){\halfwaveEllipse(1.)[D][1,0.75]}%
\mov(7,3){\halfwaveEllipse(1.)[D][1,0.75]}%
\mov(3,0){\lin(1,0)}%
\mov(0,3){\lin(4,0)}%
\mov(5,0){\lin(1,0)}%
\mov(8,0){\lin(1,0)}%
\mov(5,3){\lin(4,0)}%
\mov(3.,0.){\Circle**(0.2)}%
\mov(6.,0){\Circle**(0.2)}%
\mov(1.,0.){\Circle**(0.2)}%
\mov(8.,0.){\Circle**(0.2)}%
\mov(2.,-0.8){a}%
\mov(7.,-0.8){b}%
\mov(1.,0){\lin(2.,0)}%
\mov(6.,0){\lin(2.,0)}%
\mov(1,0){\wavelin(0,3)}%
\mov(3,0){\wavelin(0,3)}%
\mov(6,0){\wavelin(2,3)}%
\mov(8,0){\wavelin(-2,3)}%
}
\caption{Proton structure and muon self-energy effects of order
$\alpha(Z\alpha)^5$.}
\end{figure}

There exists real number of important contributions of order $\alpha^6$ which
are presented in Fig.4,5. Radiative corrections of these amplitudes including
recoil effects were studied earlier both in the Lamb shift and HFS of the
hydrogen-like systems \cite{EGS,EKS,EGS2}. Radiative photons were taken in
the Fried-Yennie (FY) gauge \cite{AA,LG,FY} where the mass shell amplitudes
don't contain infrared divergences. Infrared finiteness of the Feynman
diagrams in this gauge gives the possibility to make standard subtraction
on the mass shell without introducing the photon mass. Let us consider
radiative corrections which are determined by the self-energy insertions in
the muon line. The renormalizable mass operator in the FY gauge is equal
\cite{EGS}:
\begin{equation}
\Sigma^R(p)=\frac{\alpha}{\pi}(\hat p-m)^2\int_0^1 dx\frac{-3\hat px}
{m_1^2x+(m_1^2-p^2)(1-x)}.
\end{equation}

Making the insertion (46) in the lepton tensor of the two-photon exchange
diagrams and using the projection operators (35) we can construct the
hyperfine splitting part of the quasipotential for the amplitudes in Fig. 4.
In this case as before the vertex of the proton-photon interaction
is determined by electric and magnetic form factors because the typical loop
momenta are of order the muon mass. The contraction of the lepton and proton
tensors over the Lorentz indices and the Dirac $\gamma$ matrix trace
calculation were made in the system Form \cite{Form}. In the Euclidean space
of the variable k we can present the correction to the HFS of the muonic
hydrogen as follows:
\begin{equation}
\Delta E_{2\gamma,SE}^{HFS}=\frac{(Z\alpha)^5\mu^3}{8\pi^2}\delta_{l0}
\frac{\alpha}{\pi}\int_0^1 x dx\int_0^\infty k dk\int_0^\pi \sin^2\phi d\phi
V_{SE}(k,\phi,x),
\end{equation}
\begin{equation}
V_{SE}(k,\phi,x)=\frac{1}{(k^2+4m_2^2\cos^2\phi)[(xm_1^2+\bar xk^2)^2+
4m_1^2\bar x^2k^2\cos^2\phi]}\times
\end{equation}
\begin{displaymath}
\times\Biggl\{-\frac{4m_1^2}{m_2^2}k^2F_2^2(x+6\bar x) \cos^2\phi-\frac{8m_1^2}
{m_2^2}k^2xF_2^2+16m_1^2F_2\cos^4\phi(4F_1\bar x-F_2x-2F_2\bar x)+
\end{displaymath}
\begin{displaymath}
+16m_1^2\cos^2\phi(F_1^2x+6F_1^2\bar x+4F_1F_2x+8F_1F_2\bar x+F_2^2x+
2F_2^2\bar x)+
\end{displaymath}
\begin{displaymath}
+32m_1^2xF_1(F_1+F_2)
-\frac{4k^4}{m_2^2}F_2^2\bar x\cos^2\phi-\frac{8k^4}{m_2^2}F_2^2\bar x-
16k^2F_2^2\bar x\cos^4\phi+
\end{displaymath}
\begin{displaymath}
+16k^2\bar x\cos^2\phi(F_1^2+4F_1F_2+F_2^2)+
32k^2F_1\bar x(F_1+F_2)\Biggr\}.
\end{displaymath}
After analytical integration over the angle $\phi$ we present the
contribution (47) in the integral form which was used for numerical
calculation:
\begin{equation}
\Delta E_{2\gamma,SE}^{HFS}=E^F(2S)\frac{m_1m_2\alpha(Z\alpha)}{\pi^2(1+\kappa)}
\delta_{l0}\int_0^1 x dx\int_0^\infty dk\Biggl\{\left[-\frac{8F_2^2k^2}{m_2^2}
+32F_1(F_1+F_2)\right]\frac{1}{h_1(k,x)}+
\end{equation}
\begin{displaymath}
+\left[-\frac{k^3F_2^2}{m_2^4}-\frac{6m_1^2k^3F_2^2\bar
x}{m_2^4(xm_1^2+\bar x
k^2)}+\frac{4k}{m_2^2}(F_1^2+4F_1F_2+F_2^2)\right]
\left(\frac{1}{h_2(k,x)}-\frac{k}{h_1(k,x)}\right)+
\end{displaymath}
\begin{displaymath}
\left[\frac{2km_1^2}{m_2^2}F_2(2F_1+F_2)\bar x
-\frac{kF_2^2}{m_2^2}(xm_1^2+\bar x k^2)\right]
\left[\frac{2}{h_2^2(k,x)}-\frac{k^2}{m_2^2(xm_1^2+\bar x k^2)}\left(\frac{1}
{h_2(k,x)}-\frac{k}{h_1(k,x)}\right)\right],
\end{displaymath}
\begin{displaymath}
h_1(k,x)=k\sqrt{4m_1^2\bar x^2k^2+(xm_1^2+\bar xk^2)^2}+(xm_1^2+\bar xk^2)\sqrt
{4m_2^2+k^2},
\end{displaymath}
\begin{displaymath}
h_2(k,x)=\sqrt{4m_1^2\bar x^2k^2+(xm_1^2+\bar xk^2)^2}+(xm_1^2+\bar xk^2).
\end{displaymath}
Numerical value is equal
\begin{equation}
\Delta E_{2\gamma,SE}^{HFS}=0.0010~meV.
\end{equation}

\begin{figure}
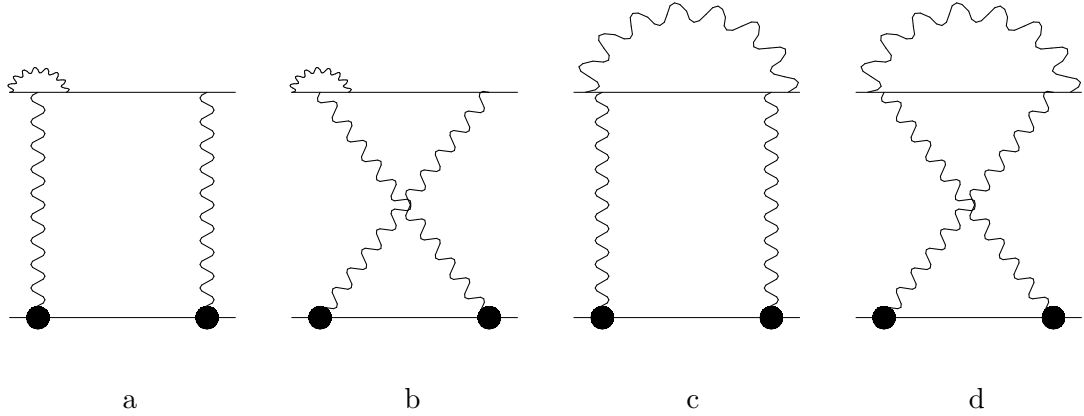

\magnitude=2000
\GRAPH(hsize=15){
\mov(0,0){\lin(2,0)}%
\mov(2.5,0){\lin(2,0)}%
\mov(2.5,2){\lin(2,0)}%
\mov(0,2){\lin(2,0)}%
\mov(5,0){\lin(2,0)}%
\mov(7.5,0){\lin(2,0)}%
\mov(7.5,2){\lin(2,0)}%
\mov(5,2){\lin(2,0)}%
\mov(0.25,0.){\Circle**(0.2)}%
\mov(0.25,0.){\wavelin(0,2)}%
\mov(1.75,0){\wavelin(0,2)}%
\mov(2.75,0){\wavelin(1.5,2)}%
\mov(4.25,0){\wavelin(-1.5,2)}%
\mov(5.25,0){\wavelin(0,2)}%
\mov(7.75,0){\wavelin(1.5,2)}%
\mov(6.75,0){\wavelin(0,2)}%
\mov(9.25,0){\wavelin(-1.5,2)}%
\mov(0.25,2){\halfwaveEllipse(0.5)[U][1.,0.8]}%
\mov(2.75,2){\halfwaveEllipse(0.5)[U][1.,0.8]}%
\mov(6.,2){\halfwaveEllipse(1.8)[U][1.,0.8]}%
\mov(8.5,2){\halfwaveEllipse(1.8)[U][1.,0.8]}%
\mov(1.75,0){\Circle**(0.2)}%
\mov(2.75,0.){\Circle**(0.2)}%
\mov(4.25,0.){\Circle**(0.2)}%
\mov(5.25,0.){\Circle**(0.2)}%
\mov(6.75,0){\Circle**(0.2)}%
\mov(7.75,0.){\Circle**(0.2)}%
\mov(9.25,0.){\Circle**(0.2)}%
\mov(1.,-0.8){a}%
\mov(3.5,-0.8){b}%
\mov(6.,-0.8){c}%
\mov(8.5,-0.8){d}%
}
\caption{Proton structure and muon vertex effects of order
$\alpha(Z\alpha)^5$.}
\end{figure}

Let us consider calculation of the vertex corrections. The renormalizable
expression of the one-particle vertex operator in the FY gauge was obtained
in Ref. \cite{ES} ($p_1^2=m_1^2$):
\begin{equation}
\Lambda_\mu^R(p,p-k)=\frac{\alpha}{4\pi}\int_0^1dx\int_0^1 dz\left[
\frac{F_\mu^{(1)}}{\Delta}+\frac{F_\mu^{(2)}}{\Delta^2}\right],
\end{equation}
where $\Delta=m_1^2x+2pk(1-x)z-k^2z(1-xz)$, the functions $F_\mu^{(1)}$, $F_\mu^{(2)}$ were
determined in Ref.\cite{ES}. The lepton tensor can be divided into two parts:
\begin{equation}
M_{\mu\nu}^{(l)(1)}=\frac{\bar v(p_1)F_\nu^{(1)}(-\hat p_1-\hat k+m_1)
\gamma_\mu v(q_1)(k^2-2k^0m_1)[m_1^2x-k^2z(1-xz)+2m_1k^2\bar x^2]}{(k^4-4k_0^2
m_1^2)[\left(m_1^2x-k^2z(1-xz)\right)^2-4m_1^2k_0^2\bar x^2z^2]},
\end{equation}
\begin{equation}
M_{\mu\nu}^{(l)(2)}=\frac{\bar v(p_1)F_\nu^{(2)}(-\hat p_1-\hat k+m_1)
\gamma_\mu v(q_1)(k^2-2k^0m_1)[m_1^2x-k^2z(1-xz)+2m_1k^2\bar x^2]^2}{(k^4-4k_0^2
m_1^2)[\left(m_1^2x-k^2z(1-xz)\right)^2-4m_1^2k_0^2\bar x^2z^2]^2}.
\end{equation}
Keeping for the simplicity the main contribution over $m_1/m_2$ we write
this type vertex corrections as follows:
\begin{equation}
\Delta E_{2\gamma,vert~1}^{HFS}=-E^F(2S)\left(\frac{\alpha}{\pi}\right)^2
\frac{8m_1m_2}{(1+\kappa)\pi}\int_0^1dx\int_0^1dz\int_0\pi\sin^2\phi
d\phi \int_0^\infty kdk\times
\end{equation}
\begin{displaymath}
\times\frac{V_1(x,k,\phi)[F_1(F_1+F_2)-(1+\kappa)]}{(k^2+4m_1^2\cos^2\phi)
(k^2+4m_2^2\cos^2\phi)\left[[m_1^2x+k^2z(1-zx)]^2+4m_1^2k^2\cos^2\phi\bar
x^2z^2\right]},
\end{displaymath}
\begin{equation}
V_1(x,k,\phi)=-2m_1^4x^2(1-x)+k^2m_1^2(6x^3z^2-8x^2z^2-3x^2z+8xz-3x)+
\end{equation}
\begin{displaymath}
+k^4(4x^3z^4-6x^2z^4-5x^2z^3+12xz^3-2xz^2-6z^2+3z),
\end{displaymath}

\begin{equation}
\Delta E_{2\gamma,vert~2}^{HFS}=-E^F(2S)\left(\frac{\alpha}{\pi}\right)^2
\frac{32m_1^3m_2}{(1+\kappa)\pi}\int_0^1x(1-x)dx\int_0^1dz\int_0\pi\sin^2
\phi d\phi\int_0^\infty k^3dk\times
\end{equation}
\begin{displaymath}
\times\frac{V_2(x,k,\phi)F_1(F_1+F_2)}{(k^2+4m_1^2\cos^2\phi)
(k^2+4m_2^2\cos^2\phi)\left[[m_1^2x+k^2z(1-zx)]^2+4m_1^2k^2\cos^2\phi\bar
x^2z^2\right]^2},
\end{displaymath}
\begin{equation}
V_2(x,k,\phi)=m_1^4x^2z(2z-1)-k^2m_1^2xz^2(4xz^2-2xz-4z+2)+
\end{equation}
\begin{displaymath}
+k^4z^3(2x^2z^3-x^2z^2-4xz^2+2xz+2z-1).
\end{displaymath}
The iteration contribution is equal
\begin{equation}
\Delta E_{iter,~2\gamma~vert}^{HFS}=<V_{1\gamma}\times G^f\times
V_{1\gamma}>_{vert}^{HFS}=F^F\left(\frac{\alpha}{\pi}\right)^2\int_0^1
dz\int_0^1dx\int_0^\infty dk\frac{4\mu}{k^2},
\end{equation}
After analytical integration in Eqs.(54) and (56) over the angle $\phi$ and
the subtraction (58) (one photon is the Coulomb-like and the other one
contains the hyperfine part of the potential with the value of magnetic form
factor at zero point) we have the expressions of the diagrams (a) and (b) in
Fig. 5:
\begin{equation}
\Delta E_{2\gamma,~vert}^{HFS}=-E^F(2S)\left(\frac{\alpha}{\pi}\right)^2
\int_0^1dx\int_0^1 dz\int_0^\infty dk\Biggl\{\frac{F_1(F_1+F_2)}{8k(1+\kappa)m_1^3m_2
\bar x^2z^2}\Bigl[-2m_1^4x^2\bar x+k^2m_1^2x\times
\end{equation}
\begin{displaymath}
\times(6x^2z^2-8xz^2-3xz+8z-3)+
k^4z(4x^3z^3-6x^2z^3-5x^2z^2+12xz^2-2xz-6z-3)\Bigr]\times
\end{displaymath}
\begin{displaymath}
\left[-\frac{\sqrt{1+b^2}}{b(a^2-b^2)(b^2-c^2)}+
\frac{\sqrt{1+a^2}}{a(a^2-b^2)(a^2-c^2)}+\frac{\sqrt{1+c^2}}{c(b^2-c^2)(a^2-c^2)}
\right]+\frac{F_1(F_1+F_2)x}{2(1+\kappa)m_1^3m_2k\bar x^3z^4}\times
\end{displaymath}
\begin{displaymath}
\left[m_1^4x^2z
(2z-1)-2k^2m_1^2xz^2(2xz^2-xz-2z+1)+k^4z^3(2x^2z^3-x^2z^2-4xz^2+2xz+2z-1)
\right]\times
\end{displaymath}
\begin{displaymath}
\times\Biggl[-\frac{\sqrt{1+b^2}}{b(a^2-b^2)(b^2-c^2)^2}+
\frac{\sqrt{1+a^2}}{a(a^2-b^2)(a^2-c^2)}+\frac{1}{2c\sqrt{1+c^2}(b^2-c^2)
(a^2-c^2)}-
\end{displaymath}
\begin{displaymath}
-\frac{\sqrt{1+c^2}}{2c^3(b^2-c^2)(a^2-c^2)}+
\frac{\sqrt{1+c^2}}{c(b^2-c^2)(a^2-c^2)}+\frac{\sqrt{1+c^2}}{c(b^2-c^2)(a^2-c^2)^2}
\Biggr]+\frac{4\mu}{k^2}\Biggr\},
\end{displaymath}
\begin{equation}
a^2=\frac{k^2}{4m_1^2},~~ b^2=\frac{k^2}{4m_2^2},~~ c^2=\frac{[m_1^2x+
k^2z(1-xz)]^2}{4m_1^2k^2\bar x^2z^2}.
\end{equation}
Numerical value of the vertex correction (59) is equal
\begin{equation}
\Delta E_{2\gamma,~vert}^{HFS}=-0.0018~meV
\end{equation}
Next vertex type diagram with one rounded photon and two exchanged photons is
the diagram of the "jellyfish" type. Its contribution to the energy spectrum
is of order  $\alpha(Z\alpha)^5$. At small loop momenta this diagram gives
the finite answer in the FY gauge. The lepton tensor relating to the diagrams
(c) and (d) in Fig.5 was obtained in Ref.\cite{EGS2}:
\begin{equation}
L_{\mu\nu}^{(\mu)}=\frac{\alpha}{4\pi}\int_0^1xdx\int_0^1 (1-z)dz\sum_{n=1}^
3\frac{M_{\mu\nu}^{(n)}}{\Delta^n},
\end{equation}
where $\Delta$ has the form as in Eq.(51). The tensor functions
$M_{\mu\nu}^{(n)}$ are written explicitly in Ref.\cite{EGS2}. The character
of further transformations of the amplitudes (c), (d) in Fig.5
to construct the HFS part of the potential is the same as for other amplitudes
shown in Figs.4,5. Omitting the details of such transformations which were
carried out by means of analytical system Form \cite{Form} we write here
three contributions to the HFS corresponding to the functions
$M_{\mu\nu}^{(n)}$ in the leading order over the ratio $m_1/m_2$:
\begin{equation}
\Delta E_{1,~jellyfish}^{HFS}=-\frac{8\alpha(Z\alpha)^5\mu^3\delta_{l0}}
{\pi^3}\int_0^1xdx\int_0^1(1-z)(1-3xz)
\int_0^\infty kdk F_1(F_1+F_2)\times
\end{equation}
\begin{displaymath}
\times\int_0^\pi\frac{\sin^2\phi d\phi}{(k^2+4m_2^2\cos^2\phi)}
\frac{[m_1^2x+k^2z(1-xz)]}{[m_1^2x+k^2z(1-xz)]^2+4m_1^2k^2\cos^2\phi
\bar x^2z^2},
\end{displaymath}
\begin{equation}
\Delta E_{2,~jellyfish}^{HFS}=-\frac{16\alpha(Z\alpha)^5\mu^3\delta_{l0}}
{3\pi^3}\int_0^1xdx\int_0^1(1-z)dz
\int_0^\infty kdk F_1(F_1+F_2)\times
\end{equation}
\begin{displaymath}
\times\int_0^\pi\frac{\sin^2\phi d\phi}{(k^2+4m_2^2\cos^2\phi)}
\frac{[m_1^2x+k^2z(1-xz)]^2[k^2xz^2(1-xz)+m_1^2(x^2z+2xz-x-3z)]}
{\left\{[m_1^2x+k^2z(1-xz)]^2+4m_1^2k^2\cos^2\phi\bar x^2z^2\right\}^2},
\end{displaymath}
\begin{equation}
\Delta E_{3,~jellyfish}^{HFS}=\frac{64\alpha(Z\alpha)^5\mu^3\delta_{l0}}
{3\pi^3}\int_0^1xdx\int_0^1(1-z)z^2dz
\int_0^\infty k^3dk m_1^2F_1(F_1+F_2)\times
\end{equation}
\begin{displaymath}
\times(x+xz-x^2z-1)\int_0^\pi\frac{\sin^2\phi d\phi}{(k^2+4m_2^2\cos^2\phi)}
\frac{[m_1^2x+k^2z(1-xz)]^3}
{\left\{[m_1^2x+k^2z(1-xz)]^2+4m_1^2k^2\cos^2\phi\bar x^2z^2\right\}^3}.
\end{displaymath}
The integration over the angle $\phi$ can be done in Eqs.(63)-(65) analytically. Omitting intermediate
expressions we can write final numerical result to the HFS of the $\mu p$:
\begin{equation}
\Delta E_{jellyfish}^{HFS}=\sum_{n=1}^3\Delta E_{n,~jellyfish}^{HFS}=
0.0005~meV.
\end{equation}
In the point-like proton approximation when the nucleus form factors entering the Feynman amplitudes
in Fig. 4,5 are changed on their values at $k^2=0$ ($F_1(0)=1$, $F_2(0)=\kappa$) the contributions
(63)-(65) will increase twofold.

\section{Conclusion}

\begin{table}
\caption{\label{t1}Corrections of orders $\alpha^5$, $\alpha^6$ to
the $2S$ state HFS in the muonic hydrogen.}
\bigskip
\begin{ruledtabular}
\begin{tabular}{|c|c|c|}  \hline
Contribution to HFS of $\mu p$ & Numerical value in meV & Reference \\   \hline
The Fermi energy $E^F(2S)$ & 22.8054  & \cite{EGS}, (12)   \\  \hline
Muon AMM correction $a_\mu E^F(2S)$ of order $\alpha^5,\alpha^6$ & 0.0266& \cite{EGS} \\ \hline
Relativistic correction $\frac{17}{8}(Z\alpha)^2 E^F(2S)$ of order $\alpha^6$ & 0.0026 &\cite{Breit}  \\  \hline
Relativistic and radiative recoil corrections&     &        \\
with the account proton AMM of order $\alpha^6$ &   0.0018 &\cite{BYG}  \\   \hline
One-loop electron vacuum polarization &    &     \\
contribution of $1\gamma$ interaction of orders $\alpha^5$, $\alpha^6$ & 0.0482& (18)  \\  \hline
One-loop muon vacuum polarization &    &   \\
contribution of $1\gamma$ interaction of order $\alpha^6$& 0.0004& (19) \\  \hline
Vacuum polarization corrections of orders $\alpha^5$, $\alpha^6$ &    &    \\
in the second order of perturbative series   & 0.0746 &(30)+(33)
\\  \hline Proton structure corrections of order $\alpha^5$ &
-0.1518 &\cite{KP},~(40) \\   \hline Proton structure corrections
of order $\alpha^6$ & -0.0017 & \cite{SGK}  \\ \hline
Electron vacuum polarization contribution+&     &     \\
proton structure corrections of order $\alpha^6$   &   -0.0026 & (43)\\  \hline
Two-loop electron vacuum polarization &     &    \\
contribution of $1\gamma$ interaction of order $\alpha^6$  & 0.0003 & (21)+(24)\\  \hline
Muon self energy + proton structure&    &    \\
correction of order $\alpha^6$   &  0.0010 & (50) \\   \hline
Vertex corrections + proton structure&    &   \\
corrections of order $\alpha^6$   &  -0.0018  & (61)\\  \hline
"Jellyfish" diagram correction + &   &    \\
proton structure corrections of order $\alpha^6$ &  0.0005  &(66) \\  \hline
HVP contribution of order $\alpha^6$ &  0.0005 & (45)\\  \hline
Proton polarizability contribution of order $\alpha^5$ & 0.0105 & \cite{CFM}\\   \hline
Weak interaction contribution & 0.0003 &  \cite{Eides} \\ \hline
Summary contribution &  22.8148 $\pm$ 0.0078 & \\   \hline
\end{tabular}
\end{ruledtabular}
\end{table}

The calculation of different quantum electrodynamical effects,
effects of the proton structure and polarizability, the hadron
vacuum polarization to the HFS of muonic hydrogen is performed in
this work. The corrections of orders $\alpha^5$ and $\alpha^6$
are considered. Working with the vacuum polarization diagrams we
take into account that the ratio $\mu\alpha/m_e$ is very close to
1 and don't increase the order of corresponding contributions.
Obtained numerical results are presented in the Table 1. We
include here also QED corrections to the Fermi energy which are
determined by the muon anomalous magnetic moment $a_\mu E^F(2S)$
\cite{EGS} (experimental value of the muon anomalous magnetic
moment $a_\mu^{exp}=11 659 203 (8)\times 10^{-10}$ \cite{Bennett}
is used), the Breit relativistic correction of order $(Z\alpha)^6$
\cite{Breit}, the proton structure corrections of order
$(Z\alpha)^6\ln (Z\alpha)^2$ \cite{SGK}, the hadron vacuum
polarization contribution \cite{FM2} and the proton
polarizability correction \cite{CFM}, the weak interaction
contribution due to the $Z$ boson exchange \cite{Eides}.

Let us point out some peculiarities of this investigation.

1.The effects of the vacuum polarization play very important role in the
case of the muonic hydrogen.
They lead to essential modification of the spin-dependent part of
the quasipotential of the one-photon interaction.

2. The proton structure effects are taken into account consistently in the loop
amplitudes by means of electromagnetic form factors. The point-like proton
approximation gives essentially increased results (approximately twofold).

3. The calculation of the muon self-energy and vertex corrections of order
$\alpha(Z\alpha)^5$ is done on the basis of the expressions for the lepton
factors in the amplitude terms of the quasipotential obtained by Eides,
Grotch and Shelyuto. We supplemented these relations by the subtraction of
the iteration terms of the potential.

Total value of the $2S$ state HFS in the muonic hydrogen shown in the Table 1
can be considered as definite guide for the future experiment which is
prepared \cite{Bakalov}. Numerical values of the corrections were obtained
with the accuracy 0.0001 meV. Theoretical error connected with the
uncertainties of fundamental physical constants (fine structure constant,
the proton magnetic moment etc.) entering the Fermi energy compose the value
near $10^{-6}$ meV. Other source of theoretical uncertainty is connected with
the corrections of higher order. Its estimation can be found from the leading
correction of the next order on $\alpha$ and $m_1/m_2$ in the form:
$\alpha(Z\alpha)^2\ln (Z\alpha)^2 E^F(2S)/\pi\approx 0.00003$ meV (the value
of fine structure constant is $\alpha^{-1}=137.03599976(50)$ \cite{MT}).
Further improvement of theoretical result presented in
the Table 1 is connected first of all with the corrections on the proton
structure and polarizability which give the theoretical error near $34\times
10^{-5}$ ppm.
The most part of this error is determined by the proton structure
corrections of order $(Z\alpha)^5$ (the Zemach correction). So, the
measurement of the hyperfine splitting of the levels $1S$ and $2S$ in
the muonic hydrogen with the accuracy $30$ ppm will lead to more accurate
value (with relative error $10^{-3}$) for the Zemach radius which than can be
used for the improvement of theoretical result for the ground state hydrogen
hyperfine structure and more reliable estimation of the proton polarizability
effect.

Another important quantity regarding to the hyperfine structure of the
muonic hydrogen is the Sternheim's hyperfine splitting interval \cite{MS0}.
It doesn't contain the proton structure and polarizability effects of the
leading order and provides additional test of quantum electrodynamics for
the hydrogen atom. Accounting numerical results of this work and
Ref.\cite{FM2004} we can write numerical value for this interval in the form:
\begin{equation}
[8\Delta E^{HFS}(2S)-\Delta E^{HFS}(1S)]=-0.120~meV.
\end{equation}
This result is valid with a precision $10^{-6}$ due to the cancellation
of the proton structure and polarizability corrections.
The increase the number of the tasks due to excited states of simple
atoms \cite{KI2002} and the inclusion new simple atoms where the hyperfine
structure of the energy spectrum is studied will decrease the
uncertainties in the determination of physical fundamental parameters and
increase the accuracy for the check of the Standard Model in low energy
physics.

\section*{Acknowledgments}

The author is grateful to D.D.Bakalov, K.Pachucki and R.N.Faustov for
fruitful discussions. This work was supported in part by the Russian
Fond for Basic Research (grant No. 04-02-16085).

\end{document}